\documentclass[aps,prl,longbibliography,twocolumn,showpacs,noeprint]{revtex4-2}

\usepackage{CJK}

\usepackage{graphicx}

\usepackage{amsmath}
\usepackage{dcolumn}

\usepackage{bm}
\usepackage[normalem]{ulem}
\usepackage{color}
\usepackage{hyperref}

\usepackage{epstopdf}
\newcommand{\bew}{\begin{widetext}}
\newcommand{\ew}{\end{widetext}}

\newcommand{\ii}{{\rm i}}
\newcommand{\bx}{\mathbf{x}}

\newcommand{\bp}{\mathbf{p}}
\newcommand{\bq}{\mathbf{q}}

\newcommand{\br}{\mathbf{r}}

\newcommand{\bff}{\mathbf{f}}

\newcommand{\bh}{\mathbf{h}}
\newcommand{\bg}{\mathbf{g}}

\newcommand{\sep}{ \ \ \ , \ \ \ }

\newcommand{\beq}{\begin{equation}}
\newcommand{\eeq}{\end{equation}}
\newcommand{\beqn}{\begin{eqnarray}}
\newcommand{\eeqn}{\end{eqnarray}}
\newcommand{\pp}{\partial}

\newcommand{\cP}{{\cal P}}

\newcommand{\vnab}{{\bf \nabla}}

\newcommand{\bG}{{\bf G}}
\newcommand{\bQ}{{\bf Q}}

\begin{document}
\title{ A new universality class describes  Vicsek’s flocking phase in physical dimensions }
\author{Patrick Jentsch}
\email{p.jentsch20@imperial.ac.uk}
\address{Department of Bioengineering, Imperial College London, South Kensington Campus, London SW7 2AZ, U.K.}
\author{Chiu Fan Lee}
\email{c.lee@imperial.ac.uk}
\address{Department of Bioengineering, Imperial College London, South Kensington Campus, London SW7 2AZ, U.K.}
\date{\today}

	\begin{abstract}
The Vicsek simulation model of flocking together with its theoretical treatment by Toner and Tu in 1995 were two foundational cornerstones of active matter physics. However, despite the field's tremendous progress, the actual universality class (UC) governing the scaling behavior of Viscek's ``flocking" phase  remains elusive. Here, we use nonperturbative, functional renormalization group methods to analyze, numerically and analytically, a simplified version of the Toner-Tu model, and uncover a novel UC with scaling exponents that agree remarkably well with the values obtained in a recent simulation study by Mahault et al.\ [Phys.~Rev.~Lett.~{\bf 123}, 218001 (2019)], in {\it  both} two and three spatial dimensions. We therefore believe that there is strong evidence that the UC uncovered here describes Vicsek's flocking phase.
	\end{abstract}

\maketitle

Two papers in 1995 arguably led to the advent of active matter physics, which has in many ways revolutionized nonequilibrium, soft matter, and biological physics: Ref. \cite{vicsek_prl95} studied the order-disorder transition of an active $XY$ model in two dimensions (2D) using a simulation model now commonly known as the Vicsek model; inspired by the appearance of an ordered  (or ``flocking") phase in 2D (forbidden by the Mermin-Wagner-Hohenberg theorem in thermal systems), Toner and Tu
introduced a set of hydrodynamic equations of motion (EOM) for generic polar active fluids in Ref.~\cite{toner_prl95}, now known as the Toner-Tu (TT) model, and investigated the scaling behavior of such a flocking phase using a renormalization group (RG) analysis. Intriguingly, controversies soon emerged regarding these two landmark studies: the {\it critical} order-disorder transition, the focus of Ref.~\cite{vicsek_prl95}, was found to be pre-empted by a discontinuous phase transition \cite{gregoire_prl04}; the RG study performed in Ref.~\cite{toner_prl95} was found to be incomplete due to neglected nonlinearities in the original analysis \cite{toner_pre12}. More recently, an extensive simulation study \cite{mahault_prl19} of  Vicsek's flocking phase  has provided estimates for the scaling exponents that deviate significantly from the original  predictions of Ref.~\cite{toner_prl95}. 
As a result, the question of {\it what universality class (UC) actually describes  Vicsek's flocking phase  remains open}. Indeed, a solution has been widely considered to be intractable using current RG methodology due to its inherent complexity \cite{toner_pre12}.

\renewcommand*{\arraystretch}{1.2}
\begin{table}
\begin{tabular}{p{0.5\linewidth} p{0.14\linewidth} p{0.14\linewidth} p{0.14\linewidth}}
\hline
Spatial dimension ($d$) & $\chi$ & $z$ & $\zeta$ \\
\hline
\hline
$d= 2:$  &&&\\
\ \ this paper & $-0.325$ & $1.325$ & $0.975$ \\
\ \ Vicsek simulation \cite{mahault_prl19} & $-0.31(2)$ & $1.33(2)$ & $0.95(2)$ \\
\ \ incompressible 
\cite{chen_natcom16,chen_a23} & ${ -0.23}$ & ${ 1.1}$ & ${ 0.67}$ \\
\ \ TT 95 / Malthusian \cite{toner_prl12} & $-0.20$ & $1.20$ & $0.6$ \\
$d= 3:$  &&&\\
\ \ this paper & $-0.65$ & $1.65$ & $0.95$ \\
\ \ Vicsek simulation \cite{mahault_prl19} & $-0.62$ & $1.77$ & $1$ \\
\ \ TT 95 / incompressible 
\cite{toner_prl95, chen_njp18} & $-0.60$ & $1.60$ & $0.8$ \\
\ \ Malthusian  \cite{chen_prl20} & $-0.45(2)$ & $1.45(2)$ & $0.73(1)$\\
\hline
\end{tabular}
\caption{Comparison of the scaling exponents obtained here [Eqs.~(\ref{cexp})], from a simulation study of the Vicsek model \cite{mahault_prl19}, from the Toner-Tu 1995 paper \cite{toner_prl95}, and from dynamic renormalization group analyses of two closely related models: the {\it incompressible} Toner-Tu model \cite{chen_natcom16,chen_njp18,chen_a23} and the {\it Malthusian} version of the Toner-Tu model \cite{toner_prl12,chen_prl20}.}
\label{tab}
\end{table}

Here, we made a significant step forward in tackling the above question using a functional renormalization group (FRG) \cite{wetterich_plb93,morris_ijopa94,ellwanger_zfpc94,berges_pr02,kopietz_b10,delamotte_b12,dupuis_pr21,canet_jopa11} analysis. Specifically, starting with a simplified version of the general TT EOM, our FRG calculation leads to a set of scaling relations that enable us to solve for the three scaling exponents: roughness exponent ($\chi$), dynamic exponent ($z$), and anisotropy exponent $(\zeta)$, which  characterize the UC of the flocking phase. 

Using the  rescaling convention,
\beq
\label{eq:rescaling}
(t,\br_\bot,x, \delta\bg, \delta\rho)\rightarrow (t e^{zl},\br_\bot e^{l},x e^{\zeta l},\delta \bg e^{\chi l}, \delta\rho e^{\chi l}) \ ,
\eeq
where, without loss of generality, the flocking direction is chosen to be along the $x$-axis,
these novel exponents are:
\begin{equation}
\label{cexp}
\chi=\frac{13(1-d)}{40} \ ,\ \  z= \frac{27+13d}{40} \ ,\ \ \zeta=\frac{41-d}{40} \ , \\
\end{equation}
for $d<11/3$ where $d$ is the spatial dimension. Remarkably, the values of these exponents agree very well with the simulation results in {\it both} two and three dimensions (falling within the given simulation errors, see Table \ref{tab}). Therefore, we believe that the new UC uncovered here describes the ordered phase of the Vicsek model.

\begin{figure*}
\includegraphics[width=\linewidth]{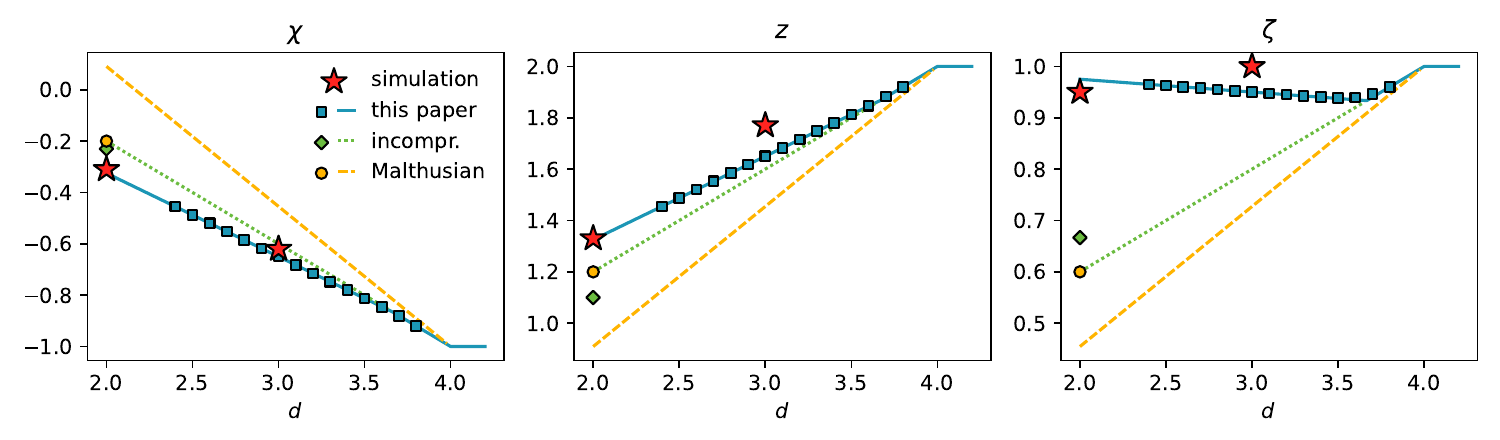}
\caption{Exponents, $\chi$, $z$ and $\zeta$, obtained from our analytical results (\ref{cexp}) (unbroken blue line) and our numeric RG calculation (blue squares) 
show good agreement with the simulation results of Vicsek's flocking phase (red stars) \cite{mahault_prl19}. For comparison, 
scaling exponents of  the incompressible TT model (green dotted line for $d>2$ \cite{chen_njp18} and green diamond for $d=2$ \cite{chen_natcom16,chen_a23}) and those of the Malthusian (or infinitely compressible) TT model (yellow dashed lines for $d>2$ \cite{chen_prl20}, yellow circle for $d=2$ \cite{toner_prl12}) are also shown. Note that the results from the Toner-Tu 1995 paper \cite{toner_prl95} coincide with the dotted green line and the yellow circle.
}
\label{fig:exponents}
\end{figure*}

{\it Simplified Toner-Tu model.---}We start with the celebrated TT EOM that describe generic compressible polar active fluids, derived simply from considering the underlying conservation law and symmetries of the system \cite{toner_prl95, toner_pre98, toner_pre12}:
\begin{align}
\label{eq:cont}
\partial_t \rho&+\nabla \cdot \bg = 0 \ ,
\\
\label{eq:mom}
\nonumber
\partial_t \bg &+\lambda_1 (\bg \cdot \nabla) \bg +{ \lambda_2\bg (\nabla \cdot \bg)+\frac{\lambda_3}{2} \vnab (|\bg|^2)} =-U\bg     \\
\nonumber
&+\mu_1 \nabla^2 \bg+ \mu_2 \nabla (\nabla \cdot \bg) + \mu_3 (\bg \cdot \nabla)^2 \bg \\
& -{ \nabla P_1}-  \bg (\bg \cdot \nabla) { P_2} +\mathrm{h.o.t.}+\bff  +\bQ \ ,
\end{align}
where $\rho$ is the mass density field and $\bg$ is the momentum density field. Note that instead of using the velocity density field as one of the two hydrodynamic variables in the original formulation \cite{toner_prl95,toner_pre98}, we have opted for the momentum field. The physics of course remains the same but this choice  has the virtue of simplifying the continuity equation (\ref{eq:cont}) by rendering it linear.
In the EOM of $\bg$ (\ref{eq:mom}), all coefficients are generic functions of $\rho$ and $|\bg|$, the ``pressure" terms $P$'s are functions of $\rho$: 
\beq
P_1 =\sum_{n\geq 1} \kappa_n (\rho-\rho_0)^n \ \ ,\ \
P_2 =\sum_{n\geq 1} \nu_n (\rho-\rho_0)^n \ ,
\eeq
where $\rho_0$ is the mean density, and  the coefficients $\kappa_n$'s and $\nu_n$'s are themselves functions of $|\bg|$.
Furthermore, ``h.o.t." in Eq.~(\ref{eq:mom}) denotes higher order  terms in  spatial derivatives (e.g., $\nabla^4 \bg$, etc) that are irrelevant to our discussion, and the noise term $\bff$ is Gaussian with vanishing mean and statistics:
\begin{equation}
\langle f_i(\br,t) f_j(\br^\prime,t^\prime)  \rangle = 2D\delta_{ij}\delta^{d+1}(t-t^\prime,\br-\br^\prime) \ .
\end{equation}

Finally, in addition to the usual terms, we have also introduced the Lagrange multiplier $\bQ$ in Eq.~(\ref{eq:mom}) to enforce that the fluctuations in $\bg$ along the flocking direction vanish ({\it Simplification 1}). While physically motivated by the fact that the $g_x$ mode is expected to be more ``massive" due to the ``potential" term $U$, we note that  our approximation here is  more drastic than the conventional nonlinear sigma constraint (as used, e.g., in the original Toner-Tu treatment \cite{toner_pre98,toner_pre12}) because instead of constraining the fluctuations on the {\it speed}, we are constraining fluctuations along the ordered direction. 

Besides {\it Simplification 1}, we will reduce the complexity further by ignoring all nonlinearities in the TT EOM involving the density field ({\it Simplification 2}). This simplification is motivated by the successes in previous studies of variants of the TT model where the density field is neglected \cite{chen_natcom16,chen_njp18,chen_a23,toner_prl12,chen_prl20}. Here, the density and momentum fields are of course still coupled at the linear level, which, as we shall see, leads to novel emergent hydrodynamic behavior.

{\it Linear Theory.---}In the flocking phase, the mean magnitude of the momentum field, $g=|\bg|$, is nonzero and we are interested in the fluctuating fields around this flocking state:
\beq
\delta \rho = \rho-\rho_0 \sep \delta \bg = \bg - g_0 \hat{\bx} \ , 
\eeq
where hats denote normalized vectors.
 We now further partition $\delta \bg$ into three components that are more natural in our analysis: $\delta \bg = \delta \bg_x +\delta \bg_L +\delta \bg_T$, where $\delta \bg_x = \hat \bx (\hat \bx\cdot \delta\bg)$, $\delta \bg_L = \hat \bq_\bot (\hat \bq_\bot\cdot \delta\bg)$, where
 $\bq_\bot$ denotes the wavevector  (in spatially transformed Fourier space)  perpendicular to the $\bx$-direction, i.e.,  
 $\bq_\bot  =\bq -q_x\hat\bx $ and $q_x = \hat \bx \cdot \bq$. Namely, the three components of $\delta \bg$ correspond to its component along the flocking direction, along the direction of the wavevector (with the $\bx$-component subtracted), and along the direction perpendicular to both wavevector and flocking direction. Note that {\it Simplification 1}  enforces that $\delta \bg_x = 0$ here.

We now analyze the scaling behavior of the ordered phase at the linear level, i.e., by first truncating the TT EOM to linear order in $\delta \rho, \delta \bg_L$, and $\delta \bg_T$. The propagators can thus be obtained by inverting the ``dynamical matrix" constructed from the linear TT EOM \cite{SI}:
\begin{align}
 \bG(\tilde \bq) &=  \bG_L(\tilde \bq)+ \bG_T(\tilde \bq) \ , \\
 \bG_L(\tilde \bq) &=  \frac{-\ii \omega_q \bm\cP_L(\bq)}{-\ii\omega_q (-\ii \omega_q+ \ii \lambda_g q_x+\mu_x^L q_x^2+\mu_\bot^L q_\bot^2) + \kappa_1 q_\bot^2} ,\\
 \bG_T(\tilde \bq) &=  \frac{\bm\cP_T(\bq)}{-\ii\omega_q + \ii \lambda_g q_x+\mu_x q_x^2+\mu_\bot q_\bot^2} \ ,
\end{align}
where we have defined $\tilde \bq =(\omega_q,\bq)$, $\cP_{L,{ij}}(\bq)= \hat q_{\bot, i}\hat q_{\bot, j}$, and $\cP_{T,{ij}}(\bq)= \delta_{ij}-\hat x_i\hat x_j-\hat q_{\bot, i}\hat q_{\bot, j}$ is the projector transverse to the $x$ and the longitudinal direction.  Further we have defined $\lambda_g = \lambda_1 g_0$, $\mu_\bot=\mu_1$, $\mu_x=\mu_1+\mu_3 g_0^2$ and $\mu_\bot^L=\mu_1+\mu_2$, which are all evaluated at $\delta\rho=\delta\bg=0$.  Since $\delta \bg_x = 0$, $\bG$ is perpendicular to $\hat {\bf x}$.

The equal-time correlation functions can then be obtained in the usual way, giving,
\begin{align}
\label{corrg}
\langle \delta\bg_L(t,\br)  \delta\bg_L(t, 0)\rangle &=  D \int_\bq e^{-\ii\bq\cdot\br} \frac{\bm{\mathcal P}_L(\bq)}{\mu_x q_x^2+\mu_\bot^L q_\bot^2}   \ ,\\
\langle \delta\bg_T(t,\br)  \delta\bg_T(t, 0)\rangle &=  D \int_\bq e^{-\ii\bq\cdot \br}\frac{\bm{\mathcal P}_T(\bq)}{\mu_x q_x^2+\mu_\bot q_\bot^2}  \ ,\\
\label{corrrho}
\langle \delta\rho(t,\br) \delta\rho(t, 0)\rangle &= \frac{D}{\kappa_1}  \int_\bq e^{-\ii\bq\cdot \br} \frac{1}{\mu_x q_x^2+\mu_\bot^L q_\bot^2}  \ ,
\end{align}
%
%
%
%
where $\int_\bq=\int d^d \bq/(2\pi)^d$. 
In particular, our linear analysis identifies the following scaling exponents $\chi^{\rm lin}_\rho=\chi^{\rm lin}=(2-d)/2, z^{\rm lin}= 2$, and $\zeta^{\rm lin}=1$, 
which, as expected, are  identical to those in previous works \cite{toner_prl95,toner_pre98,toner_pre12}.

{\it Nonlinear analysis using FRG.---} Applying {\it Simplification 2} to eliminate all nonlinearities involving $\delta \rho$, the only nonlinearities left are terms involving  the $\lambda$'s and $U$, which become independent of $\delta \rho$. 
The standard power counting method  (e.g., see \cite{forster_pra77}) shows that below $d=4$, the leading order contributions of these nonlinearities  (i.e., the $\lambda$'s, which are no longer functions of $\delta \bg$, and $U=\beta |\delta\bg|^2/2$)  can modify the scaling behavior and thus  have to be incorporated into the analysis.
RG methods provide a systematic way to accomplish this task and we will use here the functional version of the renormalization group based on the {\it exact} Wetterich equation \cite{wetterich_plb93,morris_ijopa94,ellwanger_zfpc94}:
\beq
\label{eq:wetterich}
\pp_k \Gamma_k =\frac{1}{2} {\rm Tr} \left[ \left(\Gamma^{(2)}_k +R_k\right)^{-1} \pp_k R_k\right]\ ,
\eeq
where $\Gamma_k$ is the wavelength ($k$) dependent effective average action 
 and $R_k$ is a regulator that serves to control the length scale ($\sim 1/k$) beyond which fluctuations are averaged over. The exact flow equation (\ref{eq:wetterich}) serves to interpolate between the microscopic action $\Gamma_\Lambda$ (where all model details are encoded) and the macroscopic effective average action $\Gamma_0$, from which the EOM for the {\it averages} of the fields can be obtained.
 The trace is a sum over all degrees of freedom, i.e., over all field indices, wavevectors and frequencies, and $\Gamma_k^{(2)}$ is the matrix containing the second order functional derivatives of $\Gamma_k$ with respect to the fields. The boundary conditions for $\Gamma_k$ described above are enforced by requiring that $R_\Lambda\sim \infty$ and $R_0 =0$. Otherwise, it can be chosen freely, and in principle $\Gamma_0$ is independent of that choice. In practice, we typically constrain the  form of the  microscopic action $\Gamma_\Lambda$ 
in order to close the flow equations with a finite number of coefficients, which are now also dependent on $k$.

To proceed with our NPRG analysis, we use the Martin-Siggia-Rose-de Dominicis-Janssen formalism \cite{martin_pra73,dedominicis_jpc76,janssen_zpb76,canet_jopa11}, introducing the response fields $\bar \bg_\bot$ and $\bar \rho$, to obtain a scalar action that describes our theory at the microscopic scale $\Lambda$.  Making all microscopic couplings dependent on $k$ (not written explicitly), we obtain an Ansatz for the scale-dependent effective average action, 
\begin{widetext}
\begin{align}
\label{microaction}
\nonumber
\Gamma_k[\bar\bg_\bot,\bg_\bot,\bar\rho,\rho] = &\int_{\tilde x}\left\{ \bar\rho\left[\partial_t \rho +\nabla_\bot \cdot \bg_\bot\right] - D|\bar \bg_\bot|^2+ \bar \bg_\bot \cdot \left[ \gamma\partial_t \bg_\bot +\lambda_1 g_0 \partial_x \bg_\bot +\lambda_1 (\bg_\bot \cdot \nabla_\bot) \bg_\bot + \lambda_2\bg_\bot (\nabla_\bot \cdot \bg_\bot) \right.\right. \\
\nonumber
&\hspace{0.5cm} \left.\left.+\frac{1}{2}\lambda_3 \vnab_\bot (|\bg_\bot|^2) +\frac{1}{2}\beta  |\bg_\bot|^2 \bg_\bot  - \mu_1 \nabla_\bot^2 \bg_\bot- \mu_1 \partial_x^2 \bg_\bot- \mu_2 \nabla_\bot (\nabla_\bot \cdot \bg_\bot) - \mu_3 g_0^2 \partial_x^2 \bg +\kappa_1 \nabla\rho \right] \right\} \ ,
\end{align}
\end{widetext}
 where we have defined $\bg_\bot= \delta\bg_L+\delta\bg_T$. 
We have also introduced the coefficient $\gamma$ to allow for the potential renormalization of the time-derivative term. Due to its linear structure, the ``density sector" (proportional to $\bar\rho$) does not get renormalized \cite{canet_pre15,jentsch_prr23}, and therefore its coefficients remain unity.

The last ingredient in the FRG formulation is the regulator, which we choose to be \cite{morris_npb96},
\begin{equation}
R_k(\tilde \bq,\tilde \bp)= \Gamma_k^{(2)}(\tilde \bq,\tilde \bp)\left(\frac{1}{\Theta_\epsilon(|\bq_\bot|-k)}-1\right) \ ,
\end{equation}
where $\Theta_\epsilon$ is a smooth, nonzero function that approaches the Heaviside function in the limit of $\epsilon\rightarrow 0$, which is to be taken at the end of the calculation. Note that the regulator effectively modifies the propagators with a factor independent of frequency and wavenumber in $x$-direction $q_x$. With this property, we are able to evaluate all integrals in the trace of Eq.~\eqref{eq:wetterich} analytically, except the $q_x$-integral. We further note that since the regulator has the same structure as the propagator, even though it is frequency dependent, causality is preserved \cite{canet_jopa11}.

{\it RG fixed points.---}With the regulator and ansatz defined, we can now deduce the flow equations, for which we rely on computer algebra due to the complexity of the propagators and interaction terms. Further details are given in Ref.~\cite{SI}.

Integrating the flow equations numerically, we always find a nontrivial stable fixed point. The associated scaling exponents are shown in Fig.~\ref{fig:exponents} (blue squares). For dimensions $11/3\lesssim d<4$, the scaling exponents agree with those obtained by Toner and Tu in Refs.~\cite{toner_prl95, toner_pre98}:
\begin{equation}
\label{ttexp}
\chi^{\rm TT}=\frac{3-2d}{5} \ ,\ \  z^{\rm TT}= \frac{2(d+1)}{5} \ ,\ \ \zeta^{\rm TT}=\frac{d+1}{5} \ . \\
\end{equation}
Intriguingly, below $d\approx 11/3$, the values of the exponents are found to agree with the new formula shown in 
Eq.~(\ref{cexp}), until for $d\lesssim 2.4$ when the RG flow seems to become divergent. We hypothesize that the divergence is due to our truncation/simplification of the scale-dependent average action, as we reason that the  density-dependent couplings could become more important in lower dimensions and potentially stabilize the RG flow. 

\begin{figure}
\includegraphics[width=\linewidth,trim={2.65cm 0cm 2.55cm 0cm},clip]{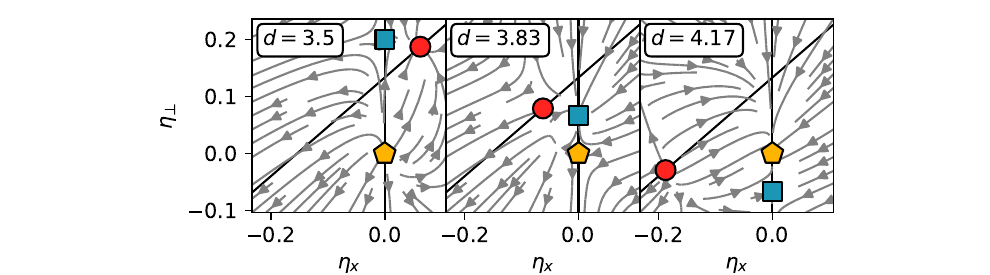}
\caption{Schematic flow diagram visualizing the locations and stability of the Gaussian (yellow pentagon), the TT (blue square) and the novel compressible fixed points (red circle) in the $(\eta_x,\eta_\bot)$-plane for dimensions below $d_c^\prime=11/3$, in between $d_c^\prime$ and $d_c=4$, and above $d_c$.
$\eta_\bot$ and $\eta_x$ are the graphical corrections of $\mu_\bot$ and $\mu_x$, respectively \eqref{eq:eta_p},\eqref{eq:eta_x}). The black lines are the fixed point trajectories as $d$ is increased. The gray flow lines are fictitious and serve to illustrate the stability of the fixed points, as verified by our numerical FRG calculation.}
\label{fig:fps}
\end{figure}

We interpret our findings as follows. The flocking phase of our simplified TT model is generically described by the TT UC for $11/3<d<4$. Below $d<11/3$, a new stable RG fixed point, with scaling  behavior given by Eq.~(\ref{cexp}), emerges. And comparing the values of the exponents obtained using Eq.~(\ref{cexp}) to a recent simulation study \cite{mahault_prl19} (red stars in Fig.~\ref{fig:exponents}), {\it we believe that the UC uncovered here describes the ordered phase of the Vicsek model.} Schematics of the RG flows illustrating the stability exchange between the TT UC and the  UC described here are shown in Fig.~\ref{fig:fps} in terms of the anomalous dimensions $\eta_\bot$ and $\eta_x$, defined as the graphical corrections of $\mu_\bot$ and $\mu_x$,
\begin{align}
\label{eq:eta_p}
\partial_l \log \mu_\bot &= z-2 +\eta_\bot \ , \\
\label{eq:eta_x}
\partial_l \log \mu_x &=z-2 \zeta +\eta_x \ .
\end{align}

{\it Analytical Treatment.---}We will now go beyond our numerical FRG calculation by using an analytical approach, whose advantage is threefold: i) to obtain analytical expressions of the scaling exponents (\ref{cexp}) beyond relying on fitting the numerical results; ii) to understand why the values of the exponents seem to be quantitatively accurate in $d=2,3$, as compared to simulations, even with drastic truncations/approximations adopted, and iii) to verify the exchange of stability between the TT UC and our new UC at $d\approx 11/3$.

To make analytical process, we start by noting that within the linear theory, the ``compressibility" term $\kappa_1 \vnab_\bot \delta \rho$ and the advective term $\lambda_g \pp_x \delta \bg \equiv  \lambda_1 g_0 \pp_x \delta \bg$ are expected to diverge as $k \rightarrow 0$, on account of their scaling dimensions based on both linear and nonlinear analyses. We can therefore approximate the flow equations by assuming the divergence of these terms (or more specifically their dimensionless versions: $\bar\kappa_1 = \kappa_1\gamma/\mu_\bot^2 k^2$ and $\bar\lambda_g=\lambda_g/\sqrt{\mu_\bot\mu_x}k$) \cite{SI}. Taking this limit, some of the graphical corrections (for $\lambda_1$, $D$ and $\lambda_g$) vanish, while others ($\lambda_2$, $\lambda_3$,  $\mu_\bot$, and $\mu_\bot^L$) become finite, but completely independent of  $\bar\kappa_1$ and $\bar\lambda_g$ \cite{SI}. 

Unexpectedly, different behaviour is observed for the graphical corrections of $\mu_x$,
\begin{align}
\label{eq:etax}
&\eta_x=
 \frac{2D}{2(d-2)\mu_x} \frac{\partial^2}{\partial q_x^2} \int_{\tilde h} \mathcal P_{T,im}(\bq) V_{ijn}(\bq,\bh) G_{na}(\tilde \bh) \\
\nonumber
&\left.\times G_{ab}(-\tilde \bh)V_{bkm}(\bq-\bh,-\bh) G_{ik}(\tilde\bq-\tilde \bh) \delta(|\bh^\bot|-k)  \right|_{\tilde\bq=0} \\
\nonumber
&\equiv F_x(\bar \lambda_1,\bar \lambda_2,\bar \lambda_3,\bar g_0, \bar \kappa_1) \ ,
\end{align}
where $\int_{\tilde h} d^d h d\omega_h/(2\pi)^{(d+1)}$ and $V_{ijn}$ is the three-point momentum density vertex (shown in Ref.~\cite{SI}). First, contrarily to the other corrections, the limit of large $\bar\kappa_1$ and $\bar\lambda_g$ does not commute with the integral over $h_x$ in Eq.~\eqref{eq:etax}. If the limit is performed before the integral, one would incorrectly conclude that this correction vanishes. Evaluated in the correct order however, we find via numerical evaluation of Eq.~\eqref{eq:etax} that, as both $\bar\kappa_1$ and $\bar \lambda_g$ diverge, and $\bar \lambda_1$, $\bar \lambda_2$ and $\bar \lambda_3$ approach a fixed point value (marked by an asterisk), the function $F_x$ behaves asymptotically as 
\begin{equation}
F_x(\bar \lambda_1,\bar \lambda_2,\bar \lambda_3,\bar \lambda_g, \bar \kappa_1) \rightarrow\bar F_x\left(\bar\lambda_1^*,\bar \lambda_2^*,\bar \lambda_3^*,\frac{\bar \lambda_g^{13}}{\bar\kappa_1^{7}}\right) \ ,
\end{equation}
where asterisks denote fixed point values and $\bar F_x$ has the asymptotic properties, $\bar F_x(\bar\lambda_1,\bar \lambda_2,\bar \lambda_3,0)=0$ and $\lim_{x\rightarrow \infty} \bar F_x(\bar\lambda_1,\bar \lambda_2,\bar \lambda_3,x) = \infty$ \cite{SI}. The flow equations therefore only allow a fixed point for two possible scenarios. Either $\bar\kappa_1$ tends to infinity fast enough such that  $\bar \lambda_g^{13}\ll\bar\kappa_1^{7}$, in which case $F_x=\eta_x=0$ and the TT UC is recovered, or $\bar\kappa_1$ and $\bar \lambda_g$ diverge in such a way that the ratio $\bar \lambda_g^{13}/\bar\kappa_1^{7}$ approaches a constant value, leading to the new UC uncovered here. Specifically,   since neither $\bar \lambda_g$ nor $\bar\kappa_1$ receive any graphical corrections in this limit, using the standard rescaling, we can write down an effective flow equation for $\bar\lambda_\kappa = \bar \lambda_g^{13}/\bar\kappa_1^{7}$:
\begin{equation}
\label{eq:flowfirst}
\partial_l \bar\lambda_\kappa = [13 (z-\zeta) - 7 (2 z- 2) ]\bar\lambda_\kappa	\ ,
\end{equation}
with precisely two fixed points: either $\bar\lambda_\kappa=0$ or $\bar\lambda_\kappa={\rm const}$. In the latter case, the exponents have to fulfill the {\it hyperscaling relation}:
\beq
13 (z-\zeta) - 7 (2 z- 2) =0 \ .
\eeq
  Together with the scaling relations from the vanishing graphical corrections  for $\lambda_1$ and $D$ (also obtained in Refs \cite{toner_prl95,toner_pre98}):
\begin{align}
\partial_l \log\lambda_1 &= z-1+\chi =0 		 \ , \\
\label{eq:slast}
\partial_l \log D &= z - 2 \chi - \zeta -  (d-1) =0 			 \ , 
\end{align}
we can determine the analytical expressions of the exponents in Eq.~(\ref{cexp}). Further, our calculation here applies below $d\sim 2.4$, thus extending our numerical FRG results to d=2.

Since the exponents are determined by the various scaling relations and therefore independent of the exact locations of the fixed points, we believe them to be robust against approximations. Indeed, {\it exact} exponents have been claimed for diverse systems based on scaling relations \cite{hwa_prl89,toner_prl95,chen_njp18,toner_prl18,mahdisoltani_prr21,chen_prl22,jentsch_a23}. Further testing the {\it exactness} of the exponents (\ref{cexp}), via simulation or numerical FRG methods, will thus be of great interest.

Finally, we can support the scenario regarding the exchange of stabilities between the TT UC and our novel UC (Fig.~\ref{fig:fps}) by  expanding Eq.~\eqref{eq:flowfirst} around the TT fixed point at $\bar\lambda_\kappa=0$, 
\begin{equation}
\partial_l \delta\lambda_\kappa = (11-3d)\ \delta\lambda_\kappa \ ,
\end{equation}
which clearly becomes unstable below $d<d_c^\prime=11/3$. Further supporting evidence is that the exponents of both TT UC \eqref{ttexp} and our UC \eqref{cexp} coincide at $d=11/3$, as expected from such an exchange of fixed point stabilities.

{\it Summary \& Outlook.---}The Vicsek model together with the Toner-Tu theoretical formulation of polar active fluids helped propel active matter physics into a well-known discipline of physics today, and along the way inspired diverse variations of flocking models that are found to correspond to many novel UCs \cite{
toner_prl12,chen_njp15,chen_natcom16,chen_njp18,chen_pre18,toner_prl18,toner_pre18,chen_prl20,chen_pre20,dicarlo_njp22,
chen_prl22,chen_pre22,chen_prl22b,zinati_pre22,jentsch_prr23,jentsch_a23,cavagna_natphys23}. 
Ironically, the UC that governs Vicsek's original flocking phase has remained unknown, perhaps until now.
Besides potentially explaining the universal flocking behavior of the Vicsek model, our nonperturbative, FRG calculation may also refute the recent questioning of the stability of the flocking phase \cite{codina_prl22,brieuc_prl23} in $d=2,3$, at least when the active systems are deep enough in the ordered phase. 

Going beyond, an immediate open question is: {\it why would Vicsek's flocking phase correspond to the simplified TT model considered here?} Another open question is
whether the ordered phase of the general TT model (without imposing our simplifications) also corresponds to the UC uncovered in this work. 
Finally, the RG has traditionally been perceived as less of a quantitative tool but more of a conceptual means to elucidating the underlying physics \cite{cardy_b96}; in contrast, our work shows that RG methodology can quantify physical properties accurately, thus demonstrating the power of the RG not readily appreciated by many physicists. Indeed, FRG methodology has already been used in several other disciplines of physics to provide quantitatively accurate predictions of {\it both} universal and nonuniversal quantities \cite{dupuis_pr21}. We believe that similar developments in active matter physics will be very fruitful.

\bibliography{references}

\end{document}


\title{ Supplemental Material to: \\ A new universality class describes  Vicsek’s flocking phase in physical dimensions }
\author{Patrick Jentsch}
\email{p.jentsch20@imperial.ac.uk}
\address{Department of Bioengineering, Imperial College London, South Kensington Campus, London SW7 2AZ, U.K.}
\author{Chiu Fan Lee}
\email{c.lee@imperial.ac.uk}
\address{Department of Bioengineering, Imperial College London, South Kensington Campus, London SW7 2AZ, U.K.}
\date{\today}

	\begin{abstract}

	\end{abstract}

\maketitle

\section{Ansatz for the effective average action}

As discussed in the main text (MT), {\it Simplification 1} enforces that $\delta \bg_x = 0$ via the Lagrange multiplier $\bQ$ in the EOM of the momentum density field $\bg$.
%
In the presence of the background field $\bg_0 = g_0 \hat \bx$ and with the expansion $\bg - g_0 \hat \bx = \bg_\bot =  \delta \bg_L+\delta \bg_T$, we can solve 
for $\bQ$ by projecting the EOM [Eq.~(3) and (4) in the MT] in the $x$-direction, to obtain,
\begin{equation}
\bQ = \hat \bx \left\{ \frac{1}{2}\lambda_3 \partial_x \left(|\delta \bg_\bot|^2\right) +  { \partial_x P_1} +   g_0^2 \partial_x { P_2} + g_0 (\bg_\bot \cdot \nabla) { P_2} - (\hat \bx\cdot\bff) \right\} \ .
\end{equation}
The remaining EOM [Eq.~(3) and (4) in the MT] can then be written as 
\begin{align}
\label{cont}
\partial_t \rho&+\nabla_\bot\cdot\bg_\bot = 0 \ , \\
\nonumber
\partial_t \bg_\bot &+\lambda_1 g_0 \partial_x \bg_\bot +\lambda_1 (\bg_\bot \cdot \nabla_\bot) \bg_\bot +{ \lambda_2\bg_\bot (\nabla_\bot \cdot \bg_\bot)+\frac{1}{2}\lambda_3 \vnab_\bot (|\bg_\bot|^2)} +U\bg_\bot  \\
&= \mu_1 \nabla_\bot^2\bg_\bot+\mu_1 \partial_x^2 \bg_\bot+ \mu_2 \nabla_\bot (\nabla_\bot \cdot \bg_\bot) + \mu_3 (g_0\partial_x + \bg_\bot \cdot \nabla_\bot)^2 \bg_\bot  -{ \nabla_\bot P_1}-  \bg_\bot (g_0\partial_x + \bg_\bot \cdot \nabla_\bot) { P_2} +\bff_\bot  \ .
\end{align}
{\it Simplification 2} now removes all nonlinearities coupling to the density, leading to,
\begin{align}
\nonumber
\partial_t \bg_\bot &+\lambda_1 g_0 \partial_x \bg_\bot +\lambda_1 (\bg_\bot \cdot \nabla_\bot) \bg_\bot +{ \lambda_2\bg_\bot (\nabla_\bot \cdot \bg_\bot)+\frac{1}{2}\lambda_3 \vnab_\bot (|\bg_\bot|^2)} +U\bg_\bot  \\
&= \mu_1 \nabla_\bot^2\bg_\bot+\mu_1 \partial_x^2 \bg_\bot+ \mu_2 \nabla_\bot (\nabla_\bot \cdot \bg_\bot) + \mu_3 (g_0\partial_x + \bg_\bot \cdot \nabla_\bot)^2 \bg_\bot  -{\kappa_1 \nabla_\bot \delta \rho}+\bff_\bot  \ .
\end{align}
where all coefficients are now taken to be independent of $\delta\rho$. As we show below, the linear scaling dimension of $\bg_\bot$ is $\chi = (2-d)/2$, and therefore the nonlinearities with the largest scaling dimension become relevant at $d_c=4$. Retaining only nonlinearities relevant just below $d=4$, the potential term $U$ becomes,
\begin{equation}
U\left(\rho_0,\frac{1}{2}|\bg_0+\bg_\bot|^2\right) = \frac{1}{2}\beta  |\bg_\bot|^2   \ ,
\end{equation}
and the EOM subsequently read,
\begin{align}
\label{eom}
\nonumber
\partial_t \bg_\bot &+\lambda_1 g_0 \partial_x \bg_\bot +\lambda_1 (\bg_\bot \cdot \nabla_\bot) \bg_\bot +{ \lambda_2\bg_\bot (\nabla_\bot \cdot \bg_\bot)+\frac{1}{2}\lambda_3 \vnab_\bot (|\bg_\bot|^2)} +\frac{1}{2}\beta  |\bg_\bot|^2 \bg_\bot  \\
&= \mu_1 \nabla_\bot^2\bg_\bot+\mu_1 \partial_x^2 \bg_\bot+ \mu_2 \nabla_\bot (\nabla_\bot \cdot \bg_\bot) + \mu_3 g_0^2 \partial_x^2 \bg_\bot  -{\kappa_1 \nabla_\bot \delta \rho}+\bff_\bot  \ ,
\end{align}
where all couplings are now independent of $\bg_\bot$ and $\delta \rho$.

Applying the Martin-Siggia-Rose-de Dominicis-Janssen (MSRDJ) formalism to Eqns.~\eqref{cont} and \eqref{eom}, we obtain a microscopic action $\Gamma_\Lambda$, to which the Functional Renormalization Group (FRG) can be applied. By making all couplings, $\lambda_1$, $\lambda_3$, $\lambda_2$, $\beta$, $\kappa_1$, $\mu_1$, $\mu_2$, $\mu_3$, and $D$ dependent on the Renormalization group scale $k$ (not written explicitly), as well as introducing a $k$ dependent factor $\gamma$ for the time derivative term, we obtain an ansatz for the scale-dependent effective average action,
\begin{align}
\label{microaction}
\nonumber
\Gamma_k[\bar\bg_\bot,\bg_\bot,\bar\rho,\rho] = &\int_{\tilde x}\left\{ \bar\rho\left[\partial_t \rho +\nabla_\bot \cdot \bg_\bot\right] - D|\bar \bg_\bot|^2+ \bar \bg_\bot \cdot \left[ \gamma\partial_t \bg_\bot +\lambda_1 g_0 \partial_x \bg_\bot +\lambda_1 (\bg_\bot \cdot \nabla_\bot) \bg_\bot + \lambda_2\bg_\bot (\nabla_\bot \cdot \bg_\bot) \right.\right. \\
\nonumber
&\hspace{0.5cm} \left.\left.+\frac{1}{2}\lambda_3 \vnab_\bot (|\bg_\bot|^2) +\frac{1}{2}\beta  |\bg_\bot|^2 \bg_\bot  - \mu_1 \nabla_\bot^2 \bg_\bot- \mu_1 \partial_x^2 \bg_\bot- \mu_2 \nabla_\bot (\nabla_\bot \cdot \bg_\bot) - \mu_3 g_0^2 \partial_x^2 \bg +\kappa_1 \nabla\rho \right] \right\} \ ,
\end{align}
where we now write $\delta\rho\rightarrow\rho$ for ease of notation.

\section{Linear analysis}
In principle, the linear analysis of the theory can be performed before taking the MSRDJ transformation, as described in the MT, however, on a technical level we find it more convenient to perform the linear analysis after the transformation. Namely, all response and correlation functions can be obtained by inverting the second-order functional derivative $\Gamma_k^{(2)}$. Since the $x$-component has been eliminated, functional derivatives with respect to $\bg_\bot$ only act in the transverse subspace. Consequently,
\begin{equation}
\label{fcon}
\frac{\delta }{\delta g_{\bot ,j}(\tilde \bq)} g_{\bot ,i}(\tilde \br) = \delta^\bot_{ij} e^{\ii \omega_q t - \ii \bq \cdot \br } \ , \ \ \frac{\delta }{\delta \bar g_{\bot ,j}(\tilde \bq)} \bar g_{\bot ,i}(\tilde \br) = \delta^\bot_{ij} e^{\ii \omega_q t - \ii \bq \cdot \br } \ , \ \ \frac{\delta }{\delta \rho(\tilde \bq)} \rho(\tilde \br) = e^{\ii \omega_q t - \ii \bq \cdot \br } \ , \ \ \frac{\delta }{\delta \bar\rho(\tilde \bq)} \bar\rho(\tilde \br) = e^{\ii \omega_q t - \ii \bq \cdot \br } \ , 
\end{equation}
where, $\tilde \br=(t,\br)$, $\tilde \bq=(\omega_q,\bq)$ and $\delta^\bot_{ij}=\delta_{ij}-\hat x_i \hat x_j$. Eq.~\eqref{fcon} also defines our Fourier Conventions. The second-order functional derivative of the effective action is therefore,
\begin{equation}
\label{eq:invprop}
    \Gamma^{(2)}(\tilde \bq,\tilde \bh)=
    \begin{pmatrix}
    -2D\bdel^\bot & \tilde \bG^{-1}(\tilde \bq) & 0 & -\bK(\tilde \bq) \\
    (\tilde \bG^{T})^{-1}(-\tilde \bq) & 0 & -\ii \bq_\bot & 0 \\
    0 & \ii \bq_\bot^T & 0 & \tilde H^{-1}(\tilde \bq) \\
    -\bK^T(-\tilde \bq) & 0 & \tilde H^{-1}(-\tilde \bq) & 0 \\
    \end{pmatrix}
    \tilde\delta_{qh} \ ,
\end{equation}
with,
\begin{align}
    \Gamma^{(0,0,1,1)}(\tilde \bq,\tilde \bh)&=\tilde H^{-1}(\tilde \bq) \tilde\delta_{qh} &=& -\ii \omega_q \tilde\delta_{qh} \ , \\ 
    \Gamma^{(0,1,1,0)}(\tilde \bq,\tilde \bh)&=&&-\ii q_{\bot, i} \tilde\delta_{qh}\ , \\ 
    \Gamma^{(2,0,0,0)}_{ij}(\tilde \bq,\tilde \bh)&=&& - 2 D \delta^\bot_{ij} \tilde\delta_{qh}\ , \\ 
\label{iprop}
    \Gamma^{(1,1,0,0)}_{ij}(\tilde \bq,\tilde \bh)&=\tilde G^{-1}_{ij}(\tilde \bq)\tilde\delta_{qh} &=&\left(-\gamma \ii \omega_q \delta^\bot_{ij} + U
    +\ii \lambda_1 g_0 \partial_x \delta^\bot_{ij}+\mu_1 q^2 \delta^\bot_{ij} + \mu_2 q_{\bot ,i} q_{\bot ,j} + \mu_3 \bg_0 q_x^2 \delta^\bot_{ij}   \right)\tilde\delta_{qh}  \ ,\\
\Gamma^{(1,0,0,1)}_i(\tilde \bq,\tilde \bh)&=-K_{i}(\tilde \bq) \tilde\delta_{qh}&=&\ii \kappa_1 q_{\bot ,i} \tilde\delta_{qh} \ ,
\end{align}
where $q_x=\hat\bx\cdot \bq$, $\bq_\bot = \bq -\hat \bx q_x$ and $\tilde\delta_{qh}=(2\pi)^{d+1}\delta^{d+1}(\tilde \bq+\tilde \bh)$. Even though $U=0$ when evaluated at the steady state, we temporarily reinstated it in Eq.~\eqref{iprop}, as a general function of an arbitrary background field $\bg_\bot=const$. This has the advantage, that we can take derivatives with respect to $g_\bot=|\bg_\bot|$ which allows us to project the graphical correction of $\beta$ from the graphical correction of the two-point function.

The inverse of Eq.~\eqref{eq:invprop} is,
\begin{align}
\label{eq:prop}
\nonumber
    &\mathcal G(\tilde \bh,\tilde \bq) = \left[\Gamma^{(2)}\right]^{-1}(\tilde \bh,\tilde \bq)=\tilde\delta_{qh}\times \\
    &\begin{pmatrix}
    0 &  \bG^T(-\tilde \bq) & 0 &  \ii  \bG^T(-\tilde \bq) \cdot \bq_\bot \tilde H(-\tilde \bq) \\
    \bG(\tilde \bq) &  2D \bG(\tilde \bq) \cdot  \bG^T(-\tilde \bq) &\bG(\tilde \bq)\cdot\bK(\tilde \bq)\tilde H(\tilde \bq)&  2\ii D  \bG(\tilde \bq) \cdot \bG^T(-\tilde \bq) \cdot \bq_\bot \tilde H(-\tilde \bq)  \\
    0 & \tilde H(-\tilde \bq) \bK^T(-\tilde \bq)\cdot \bG^T(-\tilde \bq) & 0 & H(-\tilde \bq) \\
    -\ii \tilde H(\tilde \bq)  \bq_\bot^T\cdot \bG(\tilde \bq)  &  - 2\ii D\tilde H(\tilde \bq)  \bq_\bot^T\cdot \bG(\tilde \bq) \cdot \bG^T(-\tilde \bq) &  H(\tilde \bq) &  2D\tilde H(\tilde \bq) \bq_\bot^T\cdot \bG(\tilde \bq) \cdot  \bG^T(-\tilde \bq) \cdot \bq_\bot\tilde H(-\tilde \bq) \\
    \end{pmatrix}
     \ ,
\end{align}
where we have defined,
\begin{align}
 H^{-1}(\tilde \bq) &= \tilde H^{-1}(\tilde \bq)+\ii \bq^T \cdot \tilde\bG(\tilde \bq) \cdot \bK(\tilde \bq) \ , \\
 \bG^{-1}(\tilde \bq) &=\tilde \bG^{-1}(\tilde \bq) +\ii \bK(\tilde \bq)\tilde H(\tilde \bq) \bq^T \ ,
\end{align}
which have the properties,
\begin{align}
\tilde H(\tilde \bq)\ii \bq^T &=  H(\tilde \bq) \ii \bq^T \cdot \tilde \bG(\tilde \bq) \cdot  \bG^{-1}(\tilde \bq) \ , \\
\bK^T(-\tilde \bq)  \bG^T(-\tilde \bq)  &= \tilde H^{-1}(-\tilde \bq)  H(-\tilde \bq)  \bK^T(-\tilde \bq)\cdot \tilde \bG^T(-\tilde \bq)  \ ,
\end{align}
from which the identity $ \Gamma^{(2)}(\tilde \bq,\tilde \bh) \cdot \mathcal G(\tilde \bh,\tilde \bp) =\mathrm{diag}(\bdel^\bot,\bdel^\bot,1,1)\tilde \delta_{q,-p}$ can be checked straightforwardly.

It now remains to determine the functions $ \bG(\tilde \bq)$ and $ H(\tilde \bq)$ explicitly. Decomposing $\bg_\bot=\delta \bg_L+\delta\bg_T$, i.e., into the longitudinal direction $\hat \bq_\bot$ (subscript $L$) and the transverse direction (subscript $T$), we can write,
\begin{equation}
\label{inv_prop_tilde}
 \bG^{-1}(\tilde \bq) = 
\begin{pmatrix}
 \tilde G^{-1}_{L} & 0 \\
 0& \tilde G^{-1}_{T} \bm\cP_T(\bq)
\end{pmatrix}
+\frac{1}{-\ii \omega_q}
\begin{pmatrix}
 \kappa_1 q_\bot^2 & 0 \\
 0& 0
\end{pmatrix} \ ,
\end{equation}
where $ \mathcal P_{T,ij}(\bq) =\delta_{ij}-\hat x_i \hat x_j-\hat q_{\bot,i}\hat q_{\bot,j} $, $ \mathcal P_{L,ij}(\bq) =\hat q_{\bot,i}\hat q_{\bot,j} $ and,
\begin{align}
\tilde G^{-1}_{LL}&=-\gamma\ii\omega_q + \ii \lambda_g q_x+U+\mu_x^L q_x^2+\mu_\bot^L q_\bot^2 \ ,\\
\tilde G^{-1}_{TT}&=-\gamma\ii\omega_q + \ii \lambda_g q_x+U+\mu_x q_x^2+\mu_\bot q_\bot^2 \ ,
\end{align}
with $\lambda_g=\lambda_1 g_0$, $\mu_\bot=\mu_1$, $\mu_x = \mu_1+\mu_3g_0^2$, $\mu_\bot^L=\mu_1+\mu_2$, $\mu_x^L=\mu_1+\mu_3g_0^2$. Note that $\mu_x^L=\mu_x$. There is no other tensor structure we can write down that would make both couplings independent from one another. This will also be reflected in the flow equations. 

Eq.~\eqref{inv_prop_tilde} can now be inverted,
\begin{align}
 \bG(\tilde \bq) &=   \bG_L(\tilde \bq)+ \bG_T(\tilde \bq) \ , \\
\label{propgx}
 \bG_L(\tilde \bq) & =  \bm\cP_L(\bq) G_L(\tilde \bq)=  \frac{-\ii \omega_q \bm\cP_L(\bq)}{-\ii\omega_q (-\ii\gamma \omega_q+ \ii \lambda_g q_x+U+\mu_x q_x^2+\mu_\bot^L q_\bot^2) + \kappa_1 q_\bot^2} ,\\
 \bG_T(\tilde \bq) &=  \bm\cP_T(\bq) G_T(\tilde \bq)=  \frac{\bm\cP_T(\bq)}{-\gamma\ii\omega_q + \ii \lambda_g q_x+U+\mu_x q_x^2+\mu_\bot q_\bot^2} \ ,
\end{align}
 and the density propagator is,
\begin{equation}
H(\tilde \bq) = \frac{-\ii\gamma \omega_q+ \ii \lambda_g q_x+U+\mu_x q_x^2+\mu_\bot^L q_\bot^2}{-\ii\omega_q (-\ii\gamma \omega_q+ \ii \lambda_g q_x+U+\mu_x q_x^2+\mu_\bot^L q_\bot^2) + \kappa_1 q_\bot^2} \ ,
\end{equation}
which will however not actually appear explicitly in any of the diagrams, since there are no nonlinear density terms.

The correlation functions are given in terms of $\bG$ and $\tilde H$ by the $(\bg_\bot,\bg_\bot)$ and $(\rho,\rho)$-entries of Eq.~\eqref{eq:prop}, i.e.,
\begin{align}
\langle \bg_\bot(\tilde \bq) \bg_\bot(\tilde \bp)\rangle &=  2D \bG(\tilde \bq) \cdot  \bG^T(\tilde \bp)\tilde \delta_{qp} \ ,\\
\langle \rho(\tilde \bq) \rho(\tilde \bp)\rangle &= \frac{2D}{\omega_q  \omega_p} \bq^T\cdot  \bG_L(\tilde \bq) \cdot  \bG_L( \tilde \bp) \cdot  \bp \ \tilde \delta_{qp}  \ .
\end{align}
Integrating both expressions over frequencies, and setting $U=U_0=0$ give the equal-time correlation functions,
\begin{align}
\label{corrg}
\langle \bg_\bot(t,\bq) \bg_\bot(t, \bp)\rangle &=  \frac{D}{\gamma}   \left( \frac{\bm{\mathcal P}_L(\bq)}{\mu_x q_x^2+\mu_\bot^L q_\bot^2} +\frac{\bm{\mathcal P}_T(\bq)}{\mu_x q_x^2+\mu_\bot q_\bot^2} \right) (2\pi)^d \delta^d(\bq+\bp) \ ,\\
\label{corrrho}
\langle \rho(\tilde \bq) \rho(\tilde \bp)\rangle &= \frac{D}{\kappa_1}  \frac{1}{\mu_x q_x^2+\mu_\bot^L q_\bot^2}   (2\pi)^d \delta^d(\bq+\bp)  \ ,
\end{align}
which have the same scaling behavior as the general Toner-Tu model \cite{toner_pre12}, such that we can infer $\chi^{\rm lin}_\rho=\chi^{\rm lin}=(2-d)/2$, $z^{\rm lin}= 2$, and $\zeta^{\rm lin}=1$.

\section{Flow equations}

The flow equations can be derived from the Wetterich equation in the following form, 
\begin{equation}
\label{wetterich_tilde}
\partial_l \Gamma_k = \frac{1}{2} \partial_{l^\prime} \mathrm{Tr} \log \left.\left( \Gamma_k^{(2)}+R_{k^\prime}\right)\right|_{k^\prime = k} \ ,
\end{equation}
where the two flow parameters $l=-\log(k/\Lambda)$ and $l^\prime=-\log(k^\prime/\Lambda)$ are considered independent and only set equal after the differentiation, such that it only acts on the Regulator of our choice, i.e.,
%
\begin{equation}
R_k(\tilde \bq,\tilde \bp)= \Gamma_k^{(2)}(\tilde \bq,\tilde \bp)\left(\frac{1}{\Theta_\epsilon(|\bq_\bot|-k)}-1\right) \ ,
\end{equation}
where $\Theta_\epsilon$ is a smooth, nonzero function that approaches the Heaviside function in the limit of $\epsilon\rightarrow 0$, which is to be taken at the end of the calculation. Since the regulator has precisely the same frequency structure as the 2-point function, causality is manifestly preserved.

The flow equations for all couplings can be obtained from the flow equation of the self-energy, which is obtained by applying two functional derivatives to Eq.~\eqref{wetterich_tilde},
\begin{equation}
\label{wetterich2point}
\partial_l \Gamma_k^{(ab)} =  \frac{1}{2}\partial_{l^\prime}\mathrm{Tr}\left[   \left( \Gamma_k^{(2)}+R_{k^\prime}\right)^{-1}\Gamma_k^{(2;ab)} -  \left( \Gamma_k^{(2)}+R_{k^\prime}\right)^{-1}\Gamma_k^{(2;a)}\left( \Gamma_k^{(2)}+R_{k^\prime}\right)^{-1}\Gamma_k^{(2;b)}\right]_{k^\prime = k} \ ,
\end{equation}
where $a$ and $b$ are multindices, denoting the respective field index, frequency and momentum of the corresponding derivative. 

Defining,
\begin{equation}
\bF_g =\frac{1}{VT}\left. \partial_l \Gamma_k^{(\bar \bg_\bot\bg_\bot)}\right|_{\bg_\bot=const,\rho=0} = \frac{1}{VT} \left.\frac{\delta^2\partial_l \Gamma_k}{\delta \bar \bg_\bot(\tilde \bq)\delta \bg_\bot(-\tilde \bq)}\right|_{\bg_\bot=const,\rho=0} \ ,
\end{equation}
\begin{equation}
\bF_\rho =\frac{1}{VT}\left. \partial_l \Gamma_k^{(\bar \bg_\bot\rho)}\right|_{\bg_\bot=const,\rho=0} = \frac{1}{VT} \left.\frac{\delta^2\partial_l \Gamma_k}{\delta \bar \bg_\bot(\tilde \bq)\delta \rho(-\tilde \bq)}\right|_{\bg_\bot=const,\rho=0} \ ,
\end{equation}
\begin{equation}
\bF_D =\frac{1}{VT}\left. \partial_l \Gamma_k^{(\bar \bg_\bot\bar\bg_\bot)}\right|_{\bg_\bot=const,\rho=0} = \frac{1}{VT} \left.\frac{\delta^2\partial_l \Gamma_k}{\delta \bar \bg_\bot(\tilde \bq)\delta \bar \bg_\bot(-\tilde \bq)}\right|_{\bg_\bot=const,\rho=0} \ ,
\end{equation}
where $VT=(2\pi)^{d+1}\delta^{d+1}(0)$ is the spatiotemporal volume, we can get the flow equations of each coupling via,
\begin{align}
\label{project_f}
\partial_l \kappa_1 &= \left[ \frac{\bq_\bot}{\ii q_\bot^2}\cdot \bF_\rho \right]_{\bg_\bot=0,\bq=0}	\ ,\\ 
\partial_l \lambda_g &= -\frac{\ii}{d-2} \left[ \frac{\partial}{\partial q_x} \mathrm{Tr}\bm \cP_T(\bq) \bF_g \right]_{\bg_\bot=0,\bq=0}		\ ,\\ 
\partial_l \mu_\bot &= \frac{1}{2(d-2)}  \left[ \frac{\partial^2}{\partial q_\bot^2}  \mathrm{Tr} \bm\cP_T(\bq) \bF_g \right]_{\bg_\bot=0,\bq=0}			\ ,\\ 
\partial_l \mu_x &=	 \frac{1}{2(d-2)}  \left[ \frac{\partial^2}{\partial q_x^2} \mathrm{Tr}\bm \cP_T(\bq) \bF_g \right]_{\bg_\bot=0,\bq=0}	\ ,\\ 
\partial_l \mu_\bot^L &=	 \frac{1}{2}  \left[ \frac{\partial^2}{\partial q_\bot^2} \mathrm{Tr} \bm\cP_L(\bq) \bF_g \right]_{\bg_\bot=0,\bq=0}	\ ,\\ 
\partial_l \lambda_1 &=	 -\frac{\ii}{g_0(d-2)} \left[ \frac{\partial}{\partial q_x}  \mathrm{Tr}\bm\cP_T(\bq) \bF_g \right]_{\bg_\bot=0,\bq=0}		\ ,\\
\partial_l \lambda_2 &=	- \frac{\ii}{g_0} \left[ \frac{\partial}{\partial q_\bot} \hat \bq_\bot \cdot\bF_g \cdot\hat \bx\right]_{\bg_\bot=0,\bq=0}		\ ,\\
\partial_l \lambda_3 &=	 -\frac{\ii}{g_0} \left[ \frac{\partial}{\partial q_\bot} \hat \bx\cdot \bF_g \cdot \hat \bq_\bot\right]_{\bg_\bot=0,\bq=0}		\ ,\\
\partial_l  \beta &= \frac{1}{d-2} \left[ \frac{\partial^2}{\partial g_\bot^2} \mathrm{Tr} \bm\cP_T(\bq) \bF_g \right]_{\bg_\bot=0,\bq=0} \ ,\\ 
\partial_l \gamma &=	 \frac{\ii}{d-2} \left[ \frac{\partial}{\partial \omega_q} \mathrm{Tr} \bm\cP_T(\bq) \bF_g \right]_{\bg_\bot=0,\bq=0}	\ ,\\ 
\label{project_l}
\partial_l D &=	- \frac{1}{2(d-2)} \left[  \mathrm{Tr} \bm\cP_T(\bq) \bF_D \right]_{\bq=0}	\ .
\end{align}
Note, that we can also project the flow equation of $\mu_x$ via,
\begin{equation}
\partial_l \mu_x = \frac{1}{2}  \left[ \frac{\partial^2}{\partial q_x^2} \mathrm{Tr}\bm \cP_L(\bq) \bF_g \right]_{\bq=0} \ ,
\end{equation}
which leads to exactly the same expression. This guarantees that $\mu_x=\mu_x^L$ throughout the RG flow. Since there are no density dependent nonlinearities $\bF_\rho=0$ and therefore $\kappa_1$ has no graphical correction, $\partial_l \kappa_1 = 0$.

As per Eq.~\eqref{wetterich2point}, to compute the $\bF$'s, the 3- and 4-point functions of $\Gamma_k$ are also required, which are:
\begin{align}
\nonumber
\Gamma^{(1,3,0,0)}_{ijkl}(\tilde \bq,\tilde \bh,\tilde \bk,\tilde \bl)&=\beta (\delta_{kl} \delta_{ij}+\delta_{jl}\delta_{ik} + \delta_{il} \delta_{kj} ) \tilde\delta_{qhkl} \ , \\
\Gamma^{(1,2,0,0)}_{ijk}(\tilde \bq,\tilde \bh,\tilde \bk)
    &=\left[
    -\ii \lambda_1 (k_j \delta_{ik}+ h_k \delta_{ij})
    -\ii \lambda_2 (\delta_{ik} h_j +\delta_{ij} k_k)
    -\ii \lambda_3 (k_i+h_i) \delta_{jk}\right]\tilde\delta_{qhk} \ .
\end{align}

\section{Graphical notation}

To represent the nontrivial part of the flow equations, we now adopt a graphical notation,
\begin{align}
\diagram{11} &= G_{ij}(\tilde \bq)  \ , \\
\diagram{12} &= 2 D \delta_{ij}  \ , \\
\diagram{13} &=\beta (\delta_{kl} \delta_{ij}+\delta_{jl}\delta_{ik} + \delta_{il} \delta_{kj} )  \ , \\
\nonumber
\diagram{14} &= 
    \ii \lambda_1 (q_j \delta_{ik}-h_j \delta_{ik}+ h_k \delta_{ij})
    +\ii \lambda_2 (\delta_{ik} h_j +\delta_{ij} q_k-\delta_{ij} h_k) 
    +\ii \lambda_3 q_i \delta_{jk} \\
&\equiv V_{ijk}(\bq,\bh) \ ,
\end{align}
where tildes are only written in case of frequency dependence, wavevectors are omitted in case the vertex does not depend on them, and vector indices are not written in case of contractions.

In this notation we can represent the $\bF$'s diagrammatically, justifying the term ``graphical corrections",
\begin{align}
\label{fg}
\bF_g(\tilde \bq) &= \frac{1}{2}\diagram{3} - \diagram{4} \ , \\
\bF_\rho(\tilde \bq) &= 0 \ , \\
\label{fd}
\bF_D(0) &= - \frac{1}{2} \diagram{8} \ .
\end{align}
Since there are no density-dependent nonlinearities, we obtain the same diagrams as for the incompressible limit. However, each propagator now contains both the transverse and the longitudinal mode, which is linearly coupled to the density, and each 3-vertex contains contributions from three different nonlinearities. Since the first diagram in Eq.~\eqref{fg} is wavevector and frequency independent, we have omitted writing out the wavevectors and since we only need the frequency and wavevector independent part of Eq.~\eqref{fd}, we have set them to zero already.

As for usual, wavevectors must be ``conserved" at each vertex and closed loops imply an integration over an internal wavevector and frequency, which we name $\tilde \bh=(\omega_h,\bh)$. Our choice of regulator implies, that the $\bh_\bot$-integral is constrained to the Wilsonian momentum shell by a delta function
\begin{equation}
\partial_l \Theta(|\bh_\bot|-k^\prime)|_{k=k^\prime}= \delta(|\bh_\bot|-k) k \ .
\end{equation}
The $h_x$ and $\omega_h$ integrals are unconstrained.

Hence, the graphical corrections can be written as 
\begin{equation}
\label{diag1}
\frac{1}{2}\diagram{5} = \beta Dk \int_{\tilde\bh}\delta(|\bh_\bot|-k) \left[\delta_{ij}+\frac{2}{d-1}(\delta_{ij}-\hat x_i \hat x_j)\right]  \left[(d-2) G_{T}(\tilde \bh) G_{T}(-\tilde \bh)+ G_{L}(\tilde \bh) G_{L}(-\tilde \bh)\right]  \ ,
\end{equation}
\begin{align}
\label{diag2}
\nonumber
-\diagram{6} =&2Dk\int_{\tilde\bh}\delta(|\bh_\bot|-k) \left[\lambda_1 (q_m \delta_{ik}-h_m \delta_{ik}+h_k \delta_{im})+\lambda_2 (q_k \delta_{im}-h_k \delta_{im}+h_m \delta_{ik})+\lambda_3 q_i \delta_{mk} \right]  \\
\nonumber
&\times \left[\lambda_1 (q_n \delta_{lj}-h_j \delta_{ln})+\lambda_2 (q_j \delta_{ln}-h_n \delta_{lj})+\lambda_3 (q_l-h_l) \delta_{nj} \right]   \\
\nonumber
&  \times\left[\mathcal P_{T,mn}(\bh) G_{T}(\tilde \bh) G_{T}(-\tilde \bh) +\mathcal P_{L,mn}(\bh)  G_{L}(\tilde \bh) G_{L}(-\tilde \bh) \right] \\
&\times \left[ \mathcal P_{T,kl}(\bq-\bh)  G_{T}(\tilde \bq-\tilde \bh)+\mathcal P_{L,kl}(\bq-\bh) G_{L}(\tilde \bq-\tilde \bh) \right] \ ,
\end{align}
\begin{align}
\label{diag3}
\nonumber
-\diagram{9} =&-2D^2 k \int_{\tilde\bh}\delta(|\bh_\bot|-k) \left[ \ii (\lambda_1-\lambda_2) (h_k \delta_{im}-h_m \delta_{ik})+\beta g_0 \left(\delta_{im}  \hat x_k+\delta_{mk}  \hat x_i+\delta_{ik} \hat x_m\right) \right] \\
\nonumber
&\times \left[  \ii (\lambda_1-\lambda_2) (h_n \delta_{jl}-h_l \delta_{jn})+\beta g_0 \left(\delta_{jn}  \hat x_l+\delta_{nl}  \hat x_j+\delta_{jl} \hat x_n\right)\right] \\
\nonumber
&  \times\left[\mathcal P_{T,mn}(\bh) G_{T}(\tilde \bh) G_{T}(-\tilde \bh) +\mathcal P_{L,mn}(\bh)  G_{L}(\tilde \bh) G_{L}(-\tilde \bh) \right] \\
&  \times\left[\mathcal P_{T,kl}(\bh) G_{T}(\tilde \bh) G_{T}(-\tilde \bh) +\mathcal P_{L,kl}(\bh)  G_{L}(\tilde \bh) G_{L}(-\tilde \bh) \right] \ ,
\end{align}
where 
\begin{equation}
\int_{\tilde\bh}=\int_{-\infty}^\infty \frac{ d \omega_h}{2\pi} \int_{-\infty}^\infty \frac{ d h_x}{2\pi} \int_{-\infty}^\infty \frac{ d^{d-1} \bh_\bot}{(2\pi)^{d-1}} \ .
\end{equation}

\section{Evaluation of graphical corrections}

We evaluate the frequency and angular integrals appearing in Eqns.~\eqref{diag1}-\eqref{diag3} analytically through computer algebra, as done
in our previous analysis on the multicritical point of compressible polar active fluids \cite{jentsch_prr23}. The remaining $h_x$ integral cannot be solved analytically in general and will be treated differently in the analytical and numerical approaches. The projections [Eqns.~\eqref{project_f}-\eqref{project_l}] are also calculated using computer algebra.
For the frequency integration, Cauchy's theorem can be applied straightforwardly as possible poles in the integrands are given by the poles of the propagators,
\begin{equation}
\omega_h^{(1,2)} = \frac{1}{2}\left[ \lambda_1 g_0 h_x  - \ii \left(\mu_\bot^L h_\bot^2+\mu_x h_x^2+U\right)\right] \pm \frac{1}{2}\sqrt{ 4\kappa_1 h_\bot^2+\left[\lambda_1 g_0 h_x- \ii \left(\mu_\bot^L h_\bot^2+\mu_x h_x^2+U\right] \right)^2 } \ ,
\end{equation}
\begin{equation}
\omega_h^{(3)} = \left[ \lambda_1 g_0 h_x  - \ii \left(\mu_\bot h_\bot^2+\mu_x h_x^2+U\right)\right] \ ,
\end{equation}
and the contour can be chosen such that they are always independent of external frequencies or wavevectors. Before the angular integration, it is convenient to first carry out the expansion into small external wavenumbers. As a result, the integrand will simply be a polynomial of the cosine $z$ of the angle between $\bq_\bot$ and $\bh_\bot$, $\bh_\bot\cdot \bq_\bot=|\bh_\bot||\bq_\bot| z$. Using the Jacobian for $d-1$ dimensional spherical coordinates,
\begin{equation}
\label{eq:angint}
    \int \frac{d{^{d-1} \bh_\bot}}{(2\pi)^{d-1}} =
    \frac{\Gamma(\frac{d-1}{2})}{\sqrt \pi \Gamma(\frac{d-2}{2})}\frac{S_{d-1}}{(2\pi)^{d-1}} \int_0^\infty d{ h_\bot}\, h_\bot^{d-2} \int_{-1}^1 \dd z (1-z^2)^\frac{d-4}{2} =\int_0^\infty d{h_\bot}\,  h_\bot^{d-2} \int_z \ ,
\end{equation}
where,
\begin{equation}
S_{d-1}=\frac{2\pi^{\frac{d-1}{2}}}{\Gamma\left(\frac{d-1}{2}\right)} \ ,
\end{equation}
is the surface area of the $d-1$ dimensional unit sphere, the integrals over monomials in $z$ can be determined
\begin{equation}
    \int_z z^n=\frac{\Gamma(\frac{d-1}{2})}{\sqrt \pi \Gamma(\frac{d-2}{2})} \int_{-1}^1 \dd z (1-z^2)^\frac{d-4}{2} z^{n} = \frac{\Gamma(\frac{d-1}{2})\Gamma(\frac{1+n}{2})}{\sqrt\pi \Gamma(\frac{d+n-1}{2})} \ ,
\end{equation}
for even $n$, otherwise the integral vanishes.
The integral over the modulus $h_\bot=|\bh_\bot|$ is evaluated trivially via the delta function.

Since the resulting expressions for the graphical corrections after the analytic integration are unfortunately very complicated and not very illuminating, they are presented in the supplemental Mathematica Notebook \textit{``SupplementalNotebook.nb"}.

\section{Dimensionless Couplings}

As opposed to an RG analysis around a critical point of, e.g. the Ising model, the choice of dimensionless couplings in the present problem is not immediately obvious. A naive rescaling with the running scale $k$ as the only mass-scale would correspond to a mean-field scaling of the density of $\chi_\rho=(4-d)/2$, correct for the multicritical point where $\kappa_1=\alpha_0=0$ \cite{jentsch_prr23}, but incorrect for the flocking phase, where we know from the linear analysis that mean-field scaling has to be $\chi_\rho=\chi_g=(2-d)/2$. The correct rescaling therefore has to involve the non-fine-tuned mass scale $\kappa_1$.

The correct rescaling for frequencies, wavevectors and fields can be deduced from the time-independent correlation functions [Eqns.~\eqref{corrg} and \eqref{corrrho}],
\begin{equation}
\label{resc_wavevectors}
q_\bot \rightarrow k \bar q_\bot \ ,\ \ \ \ 
q_x \rightarrow k \sqrt{\frac{\mu_\bot}{\mu_x}} \bar q_x \ ,\ \ \ \ 
\omega_q \rightarrow \frac{\mu_\bot k^2}{\gamma} \bar\omega_q \ ,
\end{equation}
\begin{equation}
\label{resc_fields}
\bg \rightarrow \sqrt{\frac{k^{d-2} D S_{d-1}}{\gamma \sqrt{\mu_\bot \mu_x} }} \tilde \bg \ ,\ \ \ \ 
\bar\bg \rightarrow \sqrt{\frac{k^{d+2}S_{d-1}\sqrt{\mu_\bot^{3}}}{ D\gamma\sqrt{\mu_x}}} \tilde{\bar \bg} \ ,\ \ \ \ 
\rho \rightarrow \sqrt{\frac{k^{d-2} DS_{d-1}}{\kappa_1 \sqrt{\mu_\bot \mu_x} }} \tilde \rho \ ,\ \ \ \ 
\bar\rho \rightarrow \sqrt{\frac{k^{d+2}\kappa_1 S_{d-1}\sqrt{\mu_\bot^{3}}}{ D\sqrt{\mu_x}}} \tilde{\bar \rho} \ ,
\end{equation}
%
%
where the factor $S_{d-1}$ was introduced for convenience, and implies the following dimensionless couplings,
%
\begin{equation}
\label{dimless1}
\bar \kappa_1 = \frac{\kappa_1\gamma}{\mu_\bot^2 k^2} \ , \ \ \ \ \bar\lambda_g=\frac{\lambda_g}{\sqrt{\mu_\bot\mu_x} k}  \ , \ \ \ \ 
\bar\mu_\bot^L=\frac{\mu_\bot^L}{\mu_\bot} \ , \ \ \ \ \bar \beta = \frac{\beta k^{d-4} D S_{d-1}}{\gamma\mu_\bot^2}\sqrt{\frac{\mu_\bot}{\mu_x}} \ ,
\end{equation}
\begin{equation}
\label{dimless3}
 \bar\lambda_1=\lambda_1\sqrt{\frac{k^{d-4}D S_{d-1}}{\gamma\mu_\bot^3} \sqrt{\frac{\mu_\bot}{\mu_x}}}  \ , \ \ \ \  \bar\lambda_2=\lambda_2\sqrt{\frac{k^{d-4}D S_{d-1}}{\gamma\mu_\bot^3} \sqrt{\frac{\mu_\bot}{\mu_x}}} \ , \ \ \ \  \bar\lambda_3=\lambda_3\sqrt{\frac{k^{d-4}D S_{d-1}}{\gamma\mu_\bot^3} \sqrt{\frac{\mu_\bot}{\mu_x}}} \ .
\end{equation}
From Eqs.~\eqref{resc_wavevectors} and \eqref{resc_fields}, we can also read off the scaling exponents,
\begin{equation}
\zeta = 1+\frac{\eta_x-\eta_\bot}{2} \ , \ \ \ \ z = 2-\eta_\bot+\eta_\gamma \ , \ \ \ \ \chi = \frac{2(2-d+\eta_D-\eta_\gamma)-\eta_\bot-\eta_x}{4} \ ,
\end{equation}
where the anomalous dimensions are defined as
\begin{align}
\partial_l \gamma &=	 \eta_\gamma \gamma	\ ,\\ 
\partial_l D &=\eta_D D	\ ,\\ 
\label{eq:eta_p}
\partial_l \bar\mu_\bot &= \eta_\bot \mu_\bot \ , \\
\label{eq:eta_x}
\partial_l \bar\mu_x &=	\eta_x \mu_x	\ .
\end{align}

The rescaled flow equations then read,
\begin{align}
\label{flowkappa}
\partial_l \bar\kappa_1 &= (2+\eta_\gamma-2\eta_\bot)\bar\kappa_1+ \frac{\gamma}{\mu_\bot^2 k^2}\partial_l \kappa_1 	\ ,\\ 
\label{flowlg}
\partial_l \bar\lambda_g &= \left(1-\frac{1}{2}\eta_\bot-\frac{1}{2}\eta_x\right)\bar\lambda_g	+\frac{1}{\sqrt{\mu_\bot\mu_x} k} \partial_l \lambda_g	\ ,\\ 
\partial_l \bar\mu_\bot^L &= -\eta_\bot \bar\mu_\bot^L	+\frac{1}{\mu_\bot}\partial_l \mu_\bot^L\ ,\\ 
\partial_l \bar\lambda_1 &=\frac{2(4-d+\eta_D-\eta_\gamma)-5\eta_\bot-\eta_x}{4}\bar\lambda_1	+\sqrt{\frac{k^{d-4}D S_{d-1}}{\gamma\mu_\bot^3} \sqrt{\frac{\mu_\bot}{\mu_x}}}\partial_l \lambda_1	\ ,\\
\partial_l \bar\lambda_3 &=\frac{2(4-d+\eta_D-\eta_\gamma)-5\eta_\bot-\eta_x}{4}\bar\lambda_3	+\sqrt{\frac{k^{d-4}D S_{d-1}}{\gamma\mu_\bot^3} \sqrt{\frac{\mu_\bot}{\mu_x}}}\partial_l \lambda_3	\ ,\\
\partial_l \bar\lambda_2 &=\frac{2(4-d+\eta_D-\eta_\gamma)-5\eta_\bot-\eta_x}{4}\bar\lambda_2  +\sqrt{\frac{k^{d-4}D S_{d-1}}{\gamma\mu_\bot^3} \sqrt{\frac{\mu_\bot}{\mu_x}}}\partial_l \lambda_2   \ ,\\
\label{lastgraphical}
\partial_l \bar\beta &=\left(4-d+\eta_D-\eta_\gamma-\frac{3}{2}\eta_\bot-\frac{1}{2}\eta_x\right)\bar\beta  +\frac{ k^{d-4} D S_{d-1}}{\gamma\mu_\bot^2}\sqrt{\frac{\mu_\bot}{\mu_x}}\partial_l \beta   \ .
\end{align}

\section*{Numerical Analysis}

\begin{figure}
\includegraphics[width=\linewidth]{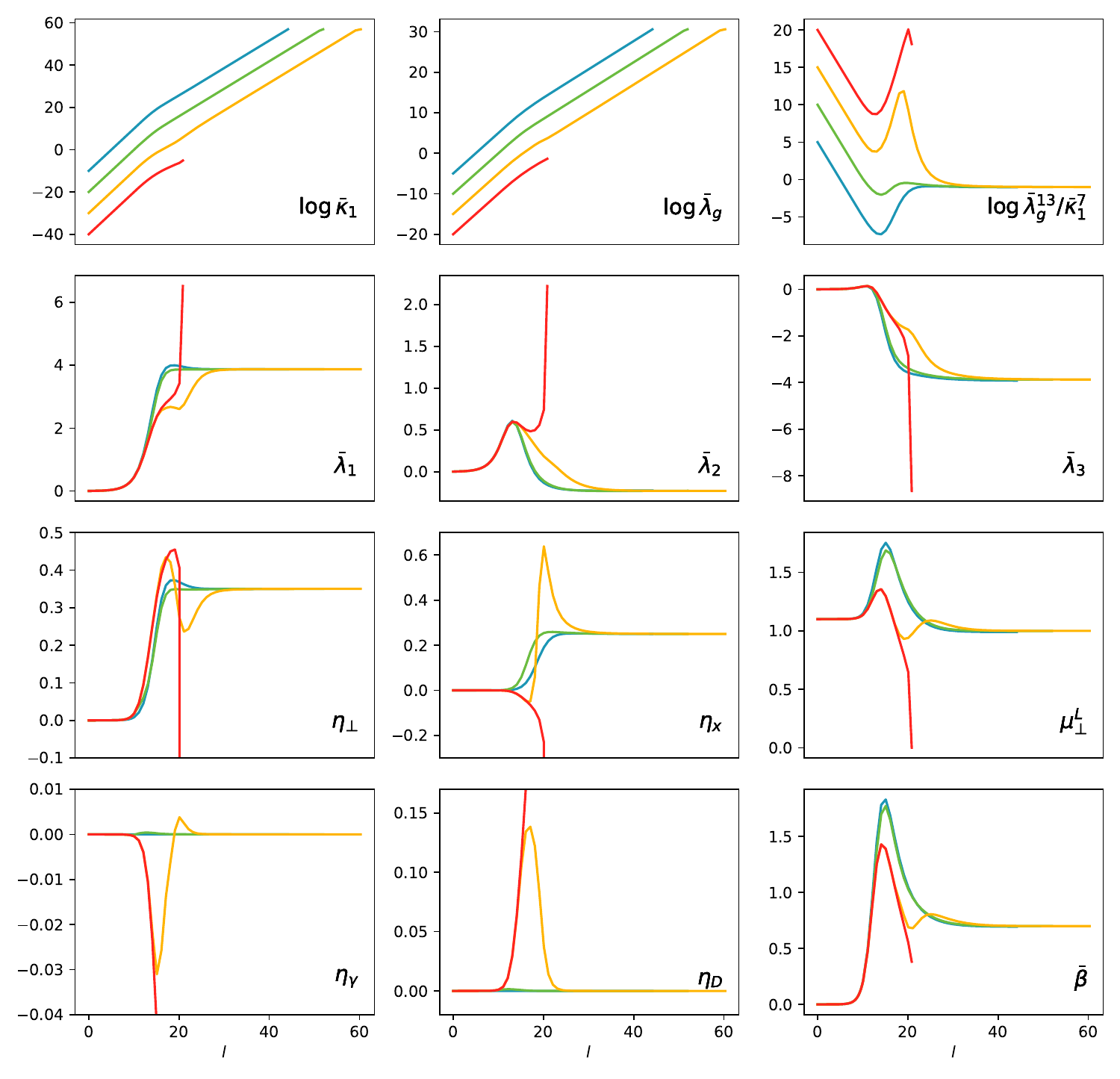}
\caption{RG flow in $d=3$ for different initial conditions of $\bar\kappa_1$. Since the initial condition for $\bar\lambda_g$ is fixed to $\bar\lambda_g=\sqrt{\bar\kappa_1}$ its initial value, as well as the initial value of the ratio $\bar\lambda_g^{13}/\bar\kappa_1^7$ vary in this plot. One can clearly see, that the numerical evaluation breaks down once $\bar\kappa_1\sim \exp(60)$ such that smaller initial values of $\bar\kappa_1$ enable longer RG-flows. However, if it is chosen too small, the RG-flow becomes divergent, the reason for which lies beyond the scope of this paper.
}
\label{couplings}
\end{figure}

For the numerical analysis of the flow equations \eqref{flowkappa}-\eqref{lastgraphical} no further approximations are made. For numerical stability, we reparametrize the flow equations as $\bar\kappa_1\rightarrow \log \bar\kappa_1$ and $\bar\lambda_g\rightarrow \bar\lambda_g/\sqrt{\bar\kappa_1}$. Then, the expressions given in \textit{``SupplementalNotebook.nb"} are converted into C++ code where the flow equations are integrated using the GNU Scientific Library's \cite{galassi_b09} implementation of a fourth-order adaptive Runge-Kutta-Fehlberg (4,5) algorithm, where at each RG ``time-step", the $h_x$ integral is solved using an adaptive quadrature routine, with the 15 point Gauss-Kronrod rule for infinite boundary integrals provided by the same library. The relative and absolute error of the ODE solver are set to $e_r = 10^{-11}$ and $e_a = 10^{-13}$ respectively and the integration errors to $e_r = 10^{-12}$ and $e_a = 10^{-10}$. This follows again closely the procedure chosen in \cite{jentsch_prr23}

The initial conditions of the flow are not important since we focus on the attractive fixed point, for which the initial values of the couplings should be forgotten exponentially. We do however perform some optimization on the initial value of $\log\bar\kappa_1$, since the flow comes to a stop when machine precision reaches its limit once $\bar\kappa_1$ becomes too large. Therefore, it must be chosen sufficiently small initially, but also not too small, since if $\kappa_1\ll 1$ at the crossover to the compressible fixed point, the RG-flow becomes unstable and diverges. Convergence to the compressible fixed point is therefore numerically constrained to a relatively small window, especially when close to the upper critical dimension $d_c=4$. This behavior is exemplified for the case $d=3$ in Fig.~\ref{couplings}. We approach this by testing different initial values of $\log\bar\kappa_1$ in a window from $-140$ to $20$ in steps of $10$ and choosing the initial conditions for which the RG-flow can be maintained the longest for our final result.

As described in the main text, we find convergence to the compressible fixed point for $2.4\lesssim d<11/3$ (note how $\eta_\bot$ and $\eta_x$ converge to $7/20$ and $1/4$ respectively in Fig.~\ref{couplings}), convergence to the incompressible fixed point for $11/3<d<4$ and convergence to the Gaussian fixed point for $4<d$. Since the values of the scaling exponents from our analytic result for $d=2$ (MT and below) match those from simulations very well, it is likely that the failure of our numeric evaluation below $d=2.4$ is a limitation of our truncation of the effective action. In fact, removing, e.g., the $\beta$ nonlinearity from our ansatz for the effective action pushes the dimension where the numerical evaluation of the flow equation fails up to $d\sim 2.5$, showing that this dimension is indeed truncation dependent.
More sophisticated truncations might be able to resolve the flow equations for smaller values of $d$.  We conjecture that these more sophisticated truncations would still be constrained by the same relations derived in our analytical approach, i.e., we believe our analytical treatment is more general than what is obtained from our numerical FRG methods.

\section*{Analytical Approach}

\begin{figure}
\includegraphics[width=\linewidth]{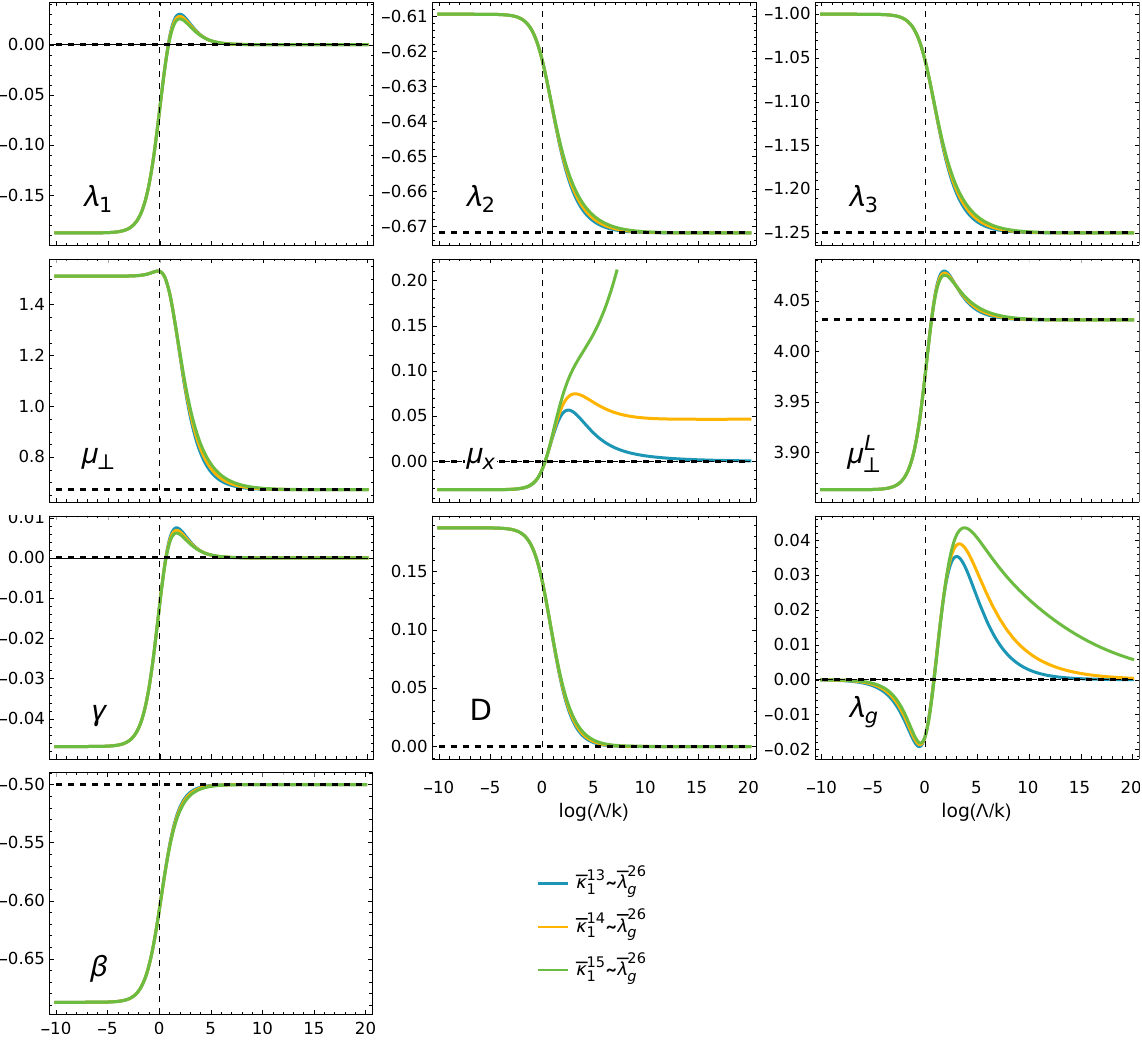}
\caption{Dimensionless graphical corrections for all couplings (indicated by watermark) evaluated at the values given by Eq.~\eqref{exvalues} and for different assumed scaling dimensions of $\bar \kappa_1$ and $\bar \lambda_g$ \eqref{assumed}. Note that this is {\it not} an RG flow. The graphical corrections of most couplings in the large $\bar \kappa_1$ and $\bar \lambda_g$ are independent of the relative scaling of the two large couplings. The coupling $\mu_x$ however behaves qualitatively very differently compared to the other couplings, which has strong implications for possible fixed point solutions. The black dashed line denotes the result obtained by taking the large $\bar \kappa_1$ and $\bar \lambda_g$ limit before performing the $h_x$ integration, which only fails for the graphical correction of $\mu_x$.
}
\label{graphcorr}
\end{figure}

Since $\bar \kappa_1$ and $\bar \lambda_g$ have positive scaling dimensions, it is expected that both diverge in the small $k$ limit. And it is typically expected that this behavior is not changed by graphical corrections. We therefore take the limit of $\bar \kappa_1,\bar \lambda_g\rightarrow \infty$. 

Since there is no UV regulation in the $h_x$ integral, this limit does not commute with the integral in general. Unfortunately, we are unable to resolve the  $h_x$ integral analytically. To get an understanding of their asymptotic behavior, we evaluate all graphical corrections at arbitrary but unequal constant values for the couplings,
\begin{equation}
\label{exvalues}
\bar\lambda_1=4 \ , \ \ \ \ \bar\lambda_2=1 \ , \ \ \ \ \bar\lambda_3=2 \ , \ \ \ \ \bar\mu_\bot^L=1  \ ,
\end{equation}
and in $d=3$ dimensions. The results that will follow are however independent of this particular choice (except for very specific choices such as $\bar\lambda_1=\bar\lambda_2$ where some graphical corrections vanish identically) and especially should also be valid at any fixed point where these couplings become constant in terms of $k$. Then we set 
\begin{equation}
\label{assumed}
\bar\kappa_1 =\left(\frac{\Lambda}{k}\right)^a \ , \ \ \ \ \ \ \bar\lambda_g =\left(\frac{\Lambda}{k}\right)^b \ ,
\end{equation}
to assume that $\bar\kappa_1$ and $\bar\lambda_g$ scale in a certain way with the RG scale $k$. Note however, that the graphical corrections do not explicitly depend on $k$, such that the physical significance encoded here is only the relative scaling of $\bar\kappa_1$ and $\bar\lambda_g$ in the limit that both are large. We then set $a$ and $b$ to particular values and plot the graphical corrections as a function of $k$ in Fig.~\ref{graphcorr}. We find that the asymptotic behavior of most of the graphical corrections is independent of the choice of $a$ and $b$ as long as they are positive such that both $\bar\kappa_1$ and $\bar\lambda_g$ grow large as $k$ becomes smaller.

The couplings $\bar\lambda_2,\bar\lambda_3,\bar\mu_\bot,\bar\mu_\bot^L$ receive nonvanishing graphical corrections that become independent of $a$, $b$, $\bar\kappa_1$ and $\bar\lambda_g$ as is generally expected. Further, the graphical correction of the couplings $\lambda_1$, $D$ and $\gamma$ will always vanish in this limit which implies the hyperscaling relationships,
\begin{equation}
\label{hyperscaling}
2(4-d+\eta_D-\eta_\gamma)-5\eta_\bot-\eta_x=0 \ , \ \ \ \ \eta_D=0 \ , \ \ \ \ \eta_\gamma = 0 \ ,
\end{equation}
at any potential fixed point as we will discuss below. 

Since the graphical correction of $\bar\lambda_g$ vanishes also, we can identify $b=1-\eta_\bot+\eta_x$, i.e., there is no additional anomalous scaling contribution to the ones we have already defined. Since $\bar\kappa_1$ receives no graphical correction at all due to the absence of nonlinear density couplings, we can also identify $a=2-2\eta_\bot+\eta_\gamma$.

Finally, the most surprising graphical correction is received by $\mu_x$, which also defines the anomalous dimension $\eta_x$. There are three different scenarios depending on the ratio $a/b$. For $a/b < 13/7$, $\eta_x$ diverges and therefore a fixed point where $a/b<13/7$ cannot exist. For $a/b>13/7$, the graphical correction of $\mu_x$ vanishes, i.e., $\eta_x=0$ in this limit. A fixed point where this is possible can exist and, together with Eqs.~\eqref{hyperscaling}, implies the same scaling exponents,
\begin{equation}
\label{expinc}
\zeta^\mathrm{TT} = \frac{d+1}{5} \ , \ \ \ \ z^\mathrm{TT} = \frac{2(d+1)}{5} \ , \ \ \ \ \chi^\mathrm{TT} = \frac{3-2d}{5} \ ,
\end{equation}
as for the TT universality class. This also fixes the scaling dimension of $\kappa_1$ and $\lambda_g$ to $a=(4d-6)/5$ and $b=(d+1)/5$ such that the actual ratio in this case is fixed to $a/b=(4d-6)/(d+1)$

There is however a different scenario in between those two cases. For exactly $a/b=13/7$, $\eta_x$ can take a constant value. Since the scaling exponents $a$ and $b$ arise from $\eta_x$ though, this value cannot be arbitrary. To be self-consistent, $\eta_x$ has to fulfill the unprecedented hyperscaling relationship,
\begin{equation}
\frac{a}{b} = \frac{ 2-2\eta_\bot+\eta_\gamma }{1+\frac{\eta_x-\eta_\bot}{2}} = \frac{ 2-2\eta_\bot+\eta_\gamma }{\zeta} = \frac{13}{7} \ .
\end{equation}
Surprisingly, together with Eq.~\eqref{hyperscaling} all anomalous dimensions are then fixed to,
\begin{equation}
\eta_x^\mathrm{}= \frac{11-3d}{8}  \ ,\ \ \eta_\bot^\mathrm{} = \frac{53-13d}{40} \ ,
\end{equation}
and imply the scaling exponents,
\begin{equation}
\label{expcompressible}
\chi^\mathrm{}=\frac{13(1-d)}{40} \ ,\ \  z^\mathrm{}= \frac{27+13d}{40} \ ,\ \ \zeta^\mathrm{}=\frac{41-d}{40} \ ,
\end{equation}
i.e., the main results of this work.

\section*{Stability of the fixed points}

For this scaling behavior to be realized the corresponding fixed point would also have to be stable. To assess the stability of the fixed point we can write down effective flow equations for the large $\bar\kappa_1$ and $\bar\lambda_g$ limit that are informed by the previous analysis. We first try to find analytic expressions for the graphical corrections by commuting the large $\bar\kappa_1$ and $\bar\lambda_g$ limit with the $h_x$ integration and comparing with the result obtained from the numerical solution of these integrals (Fig.~\ref{graphcorr}). Where we obtain the same result, we assume the limit can be commuted with the integral and we are able to solve the $h_x$-integral analytically, leading to the much simpler flow equations,
\begin{align}
\label{flowf}
\partial_l \bar\mu_\bot^L &= -\eta_\bot \bar\mu_\bot^L+ \frac{1}{32(d^2-1)(\bar\mu_\bot^L)^{\frac{3}{2}}}\left[(d+1)(\bar\lambda_1+\bar\lambda_3)^2+2(4d^2-13)(\bar\lambda_1+\bar\lambda_3)\bar\lambda_2-(8d^2-31d+27)\bar\lambda_2^2\right.	\\
\nonumber
&\hspace{4cm}\left.+(d-2)(\bar\mu_\bot^L)^{\frac{3}{2}}\left(2(4d-7)\bar\lambda_1^2+4(d^2-1)\bar\lambda_3\bar\lambda_2+(d+1)(\bar\lambda_3+8\bar\lambda_2)\bar\lambda_1\right) \right]\ ,\\ 
\partial_l \bar\lambda_1 &=\frac{2(4-d+\eta_D-\eta_\gamma)-5\eta_\bot-\eta_x}{4}\bar\lambda_1	\ ,\\
\partial_l \bar\lambda_2 &=\frac{2(4-d+\eta_D-\eta_\gamma)-5\eta_\bot-\eta_x}{4}\bar\lambda_2  -\frac{\bar\beta}{32} \left(\frac{(d-2)\bar\lambda_1}{d-1}+4(d-2)\bar\lambda_3+\frac{\bar\lambda_1+\bar\lambda_3+(8d-15)\bar\lambda_2}{(d-1)(\bar\mu_\bot^L)^{\frac{3}{2}}}\right)      \ , \\
\partial_l \bar\lambda_3 &=\frac{2(4-d+\eta_D-\eta_\gamma)-5\eta_\bot-\eta_x}{4}\bar\lambda_3	- \frac{ \bar\beta}{8} \left(\frac{2(d-2)\bar\lambda_1}{d-1}+(d-2)\bar\lambda_3+\frac{(d-1)(\bar\lambda_1+\bar\lambda_3)+(3-d)\bar\lambda_2}{(d-1)(\bar\mu_\bot^L)^{\frac{3}{2}}} \right)	\ ,\\
\partial_l \bar\beta &=\left(4-d+\eta_D-\eta_\gamma-\frac{3}{2}\eta_\bot-\frac{1}{2}\eta_x\right)\bar\beta  -(d+1)\frac{\bar\beta^2}{8} \left( \frac{1}{ (\bar\mu_\bot^L)^{\frac{3}{2}}} +(d-2) \right)  \ , \\
\eta_D &= 0 \ , \\
\eta_\gamma &= 0 \ , \\
\eta_\bot &=  \frac{(3-6d+2d^2)(\bar\mu_\bot^L)^{\frac{3}{2}}\bar\lambda_1^2+(2d-1)(\bar\lambda_1+\bar\lambda_3)\bar\lambda_2+(2d-9)\bar\lambda_2^2}{16(d^2-1)(\bar\mu_\bot^L)^{\frac{3}{2}}}\ , \\
\label{flowl}
\eta_x &= F_x(\bar \lambda_1,\bar \lambda_2,\bar \lambda_3,\bar g_0, \bar \kappa_1) \ , 
\end{align}
where we wrote the graphical correction of $\mu_x$ as $F_x$, for which we don't know the analytic expression but from our previous analysis we know it has the property,
\begin{equation}
F_x(\bar \lambda_1,\bar \lambda_2,\bar \lambda_3,\bar \lambda_g, \bar \kappa_1) \rightarrow \bar F_x\left(\bar \lambda_1^*,\bar \lambda_2^*,\bar \lambda_3^*,\frac{\bar \lambda_g^{13}}{\bar\kappa_1^{7}}\right) \ ,
\end{equation}
in the limit of large $\bar\kappa_1$ and $\bar\lambda_g$ and sufficiently close to a fixed point (indicated by the asterisk). $\bar F_x$ has the asymptotic properties, $\bar F_x(\lambda_1^*,\bar \lambda_2^*,\bar \lambda_3^*,0)=0$ and $\lim_{x\rightarrow \infty} \bar F_x(\lambda_1^*,\bar \lambda_2^*,\bar \lambda_3^*,x) = \infty$, as shown in Fig.~\ref{graphcorr}. 

\begin{figure}
\includegraphics[width=\linewidth,trim={2.65cm 0cm 2.55cm 0cm},clip]{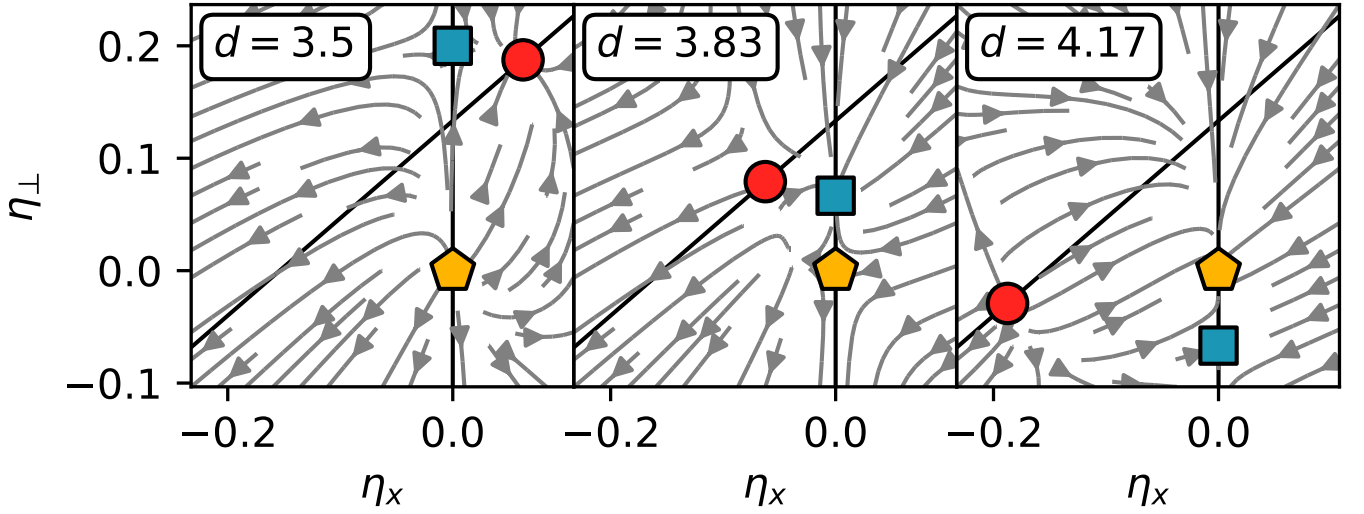}
\caption{ Schematic flow diagram visualizing the locations and stabilities of the Gaussian (yellow pentagon), the TT (blue square) and the novel compressible fixed points (red circle) in the $(\eta_x,\eta_\bot)$-plane for dimensions above $d_c  =4$, in between $d_c$ and $d_c^\prime  = 11/3$ and below $d_c^\prime$. 
$\eta_\bot$ and $\eta_x$ are the graphical corrections of $\mu_\bot$ and $\mu_x$, respectively [Eqns.~\eqref{eq:eta_p} and \eqref{eq:eta_x}]. The black lines symbolize the fixed point trajectories as $d$ is lowered. The gray flow lines are fictitious and serve to illustrate the stability of the fixed points, as verified by our numerical FRG calculation.}
\label{fig:fps}
\end{figure}

In the limit of large $\bar\kappa_1$ and $\bar\lambda_g$, the flow equations therefore cease to depend on both of these couplings but only depend on the specific combination $\bar\lambda_\kappa={\bar \lambda_g^{13}}/{\bar\kappa_1^{7}}$, which is expected to have a fixed point at either $\bar\lambda_\kappa=0$, where $\eta_x=0$, or at a finite fixed point value. We therefore combine Eqns.~\eqref{flowkappa} and \eqref{flowlg} into the additional effective flow equation
\begin{equation}
\label{auxflow}
 \partial_l\bar\lambda_\kappa =  \left[-1+ \frac{15}{2}\eta_\bot-\frac{13}{2}\eta_x-7\eta_\gamma\right]\bar\lambda_\kappa = [13 (z-\zeta) - 7 (2 z- 2) ]]\bar\lambda_\kappa \ .
\end{equation}
Since we do not have an analytic form for all of the flow equations, we cannot assess the stability of the new fixed point via our analytical approach. However, we can assess the stability of the incompressible fixed point, for which we know that $\bar\lambda_\kappa=0$, i.e., using the exponents in Eq.~\eqref{expinc}, corresponding to $\eta_\gamma=\eta_D=\eta_x=0$ and  $\eta_\bot = 2(4-d)/5$, Eq.~\eqref{auxflow} becomes,
\begin{equation}
\partial_l\bar\lambda_\kappa =  (11-3d)\bar\lambda_\kappa \ ,
\end{equation}
suggesting that the fixed point is unstable against a perturbation of $\bar\lambda_\kappa$ below a critical dimension $d^\prime_c=11/3$. At this dimension, the exponents of the incompressible fixed point coincide with those of the compressible fixed point [Eqns.~\eqref{expinc} and \eqref{expcompressible}], suggesting these fixed points collide at this critical dimension. Since fixed points usually exchange stability upon collision, it is reasonable to assume that the compressible fixed point becomes stable below $d^\prime_c=11/3$, as observed in the previous numerical analysis of the flow equation. This is illustrated in Fig.~\ref{fig:fps}.

This analysis confirms the result of the numerical evaluation of the flow equations.



\bibliography{references}